\documentclass[journal]{IEEEtran}

\usepackage{xcolor}

\usepackage[switch, pagewise]{lineno}
\usepackage{orcidlink}

\usepackage{subcaption}

\usepackage{hyperref}

\usepackage[symbol]{footmisc}

\usepackage{algorithmic}
\usepackage{graphicx}
\usepackage{amsmath}
\usepackage{textcomp}
\usepackage{xcolor}
\def\BibTeX{{\rm B\kern-.05em{\sc i\kern-.025em b}\kern-.08em
    T\kern-.1667em\lower.7ex\hbox{E}\kern-.125emX}}

\usepackage{adjustbox}
\usepackage{multirow}
\usepackage{amssymb,mathtools}

\usepackage{array}
\newcolumntype{P}[1]{>{\centering\arraybackslash}p{#1}}
\newcolumntype{M}[1]{>{\centering\arraybackslash}m{#1}}

\usepackage{tikz}
\usepackage{xcolor}
\usepackage[linesnumbered,ruled,vlined]{algorithm2e}
\SetCommentSty{mycommfont}

\SetKwInput{KwInput}{Input}                % Set the Input
\SetKwInput{KwOutput}{Output}              % set the Output

\usepackage{graphicx}

\ifCLASSINFOpdf
\else
\fi
\begin{document}
%\linenumbers

\title{AIDPS:Adaptive Intrusion Detection and Prevention System for Underwater Acoustic Sensor Networks} 

\author{Soumadeep Das \orcidlink{0009-0000-7756-6676},
        Aryan Mohammadi Pasikhani,~\IEEEmembership{Member,~IEEE}, 
        Prosanta Gope,~\IEEEmembership{Senior Member,~IEEE,}
     John A. Clark,
     Chintan Patel ~\IEEEmembership{Member,~IEEE}, and Biplab Sikdar,~\IEEEmembership{Senior Member,~IEEE}% <-this % stops a space
% \thanks{M. Shell was with the Department
% of Electrical and Computer Engineering, Georgia Institute of Technology, Atlanta,
% GA, 30332 USA e-mail: (see http://www.michaelshell.org/contact.html).}% <-this % stops a space
% \thanks{J. Doe and J. Doe are with Anonymous University.}% <-this % stops a space
% \thanks{Manuscript received April 19, 2005; revised August 26, 2015.}

\IEEEcompsocitemizethanks{\IEEEcompsocthanksitem S. Das, A. Pasikhani, P. Gope, J. Clark, C. Patel are with Department of Computer Science, University of Sheffield, Regent Court, Sheffield S1 4DP, United Kingdom.
(E-mail: [s.das2, a.mohammadipasikhani, p.gope, john.Clark, c.j.patel]@sheffield.ac.uk).
B. Sikdar is from the National University of Singapore (E-mail: bsikdar@nus.edu.sg).
%%% note need leading \protect in front of \\ to get a newline within \thanks as
%%% \\ is fragile and will error, could use \hfil\break instead.

%\IEEEcompsocthanksitem
% S. Li is with Department of Computer Science, Cardiff University, Cardiff CF10 3AT. (Email: lis117@cardiff.ac.uk)

% This work was supported
% in part by the Engineering and Physical Sciences Research Council (EPSRC) underAward EP/V039156/1. 

\textbf{Corresponding author:} Dr. Prosanta Gope
%Corresponding author: Prosanta Gope
}
}

\markboth{IEEE/ACM TRANSACTIONS ON NETWORKING}%
{Shell \MakeLowercase{\textit{et al.}}: Bare Demo of IEEEtran.cls for IEEE Journals}



\maketitle

\begin{abstract}

Underwater Acoustic Sensor Networks (UW-ASNs) are predominantly used for underwater environments and find applications in many areas. However, a lack of security considerations, the unstable and challenging nature of the underwater environment, and the resource-constrained nature of the sensor nodes used for UW-ASNs (which makes them incapable of adopting security primitives) make the UW-ASN prone to vulnerabilities. This paper proposes an Adaptive decentralised Intrusion Detection and Prevention System called AIDPS for UW-ASNs. The proposed AIDPS can improve the security of the UW-ASNs so that they can efficiently detect underwater-related attacks (e.g., blackhole, grayhole and flooding attacks). To determine the most effective configuration of the proposed construction, we conduct a number of experiments using several state-of-the-art machine learning algorithms (e.g., Adaptive Random Forest (ARF), light gradient-boosting machine, and  K-nearest neighbours) and concept drift detection algorithms (e.g., ADWIN, kdqTree, and Page-Hinkley). 
Our experimental results show that incremental ARF using ADWIN provides optimal performance when implemented with One-class support vector machine (SVM) anomaly-based detectors. Furthermore, our extensive evaluation results also show that the proposed scheme outperforms state-of-the-art bench-marking methods while providing a wider range of desirable features such as scalability and complexity.

\end{abstract}

\begin{IEEEkeywords}
Underwater Acoustic Sensor Networks, Intrusion Detection System, Incremental Machine Learning, Concept-drift Detection.
\end{IEEEkeywords}

% \fi
\IEEEpeerreviewmaketitle

\section{Introduction}
\label{sec:Introduction}

Water covers more than 70\% of the earth's surface and is also home to many natural resources. Most of these natural resources are inaccessible and unexplored. Hence, many countries have invested in monitoring and analysing sensing data observed from underwater environments (deep and shallow water) \cite{berlian2016design}. In this regard, Underwater Wireless Acoustic Sensor Networks (UW-ASNs) are an emerging technology for underwater exploration \cite{felemban2015underwater}. UW-ASNs have various applications, namely, habitat and natural resource exploration, border surveillance, disaster forecasting, navigation control, and safety-and-control. The components of UW-ASNs comprise of several sensor nodes, underwater sink, surface station, and surface sink (a.k.a. buoy) \cite{AKYILDIZ2005257}. Each component coordinates and shares information to carry out their tasks.

\begin{figure*}[htp]
    \centering
    \includegraphics[width=18cm]{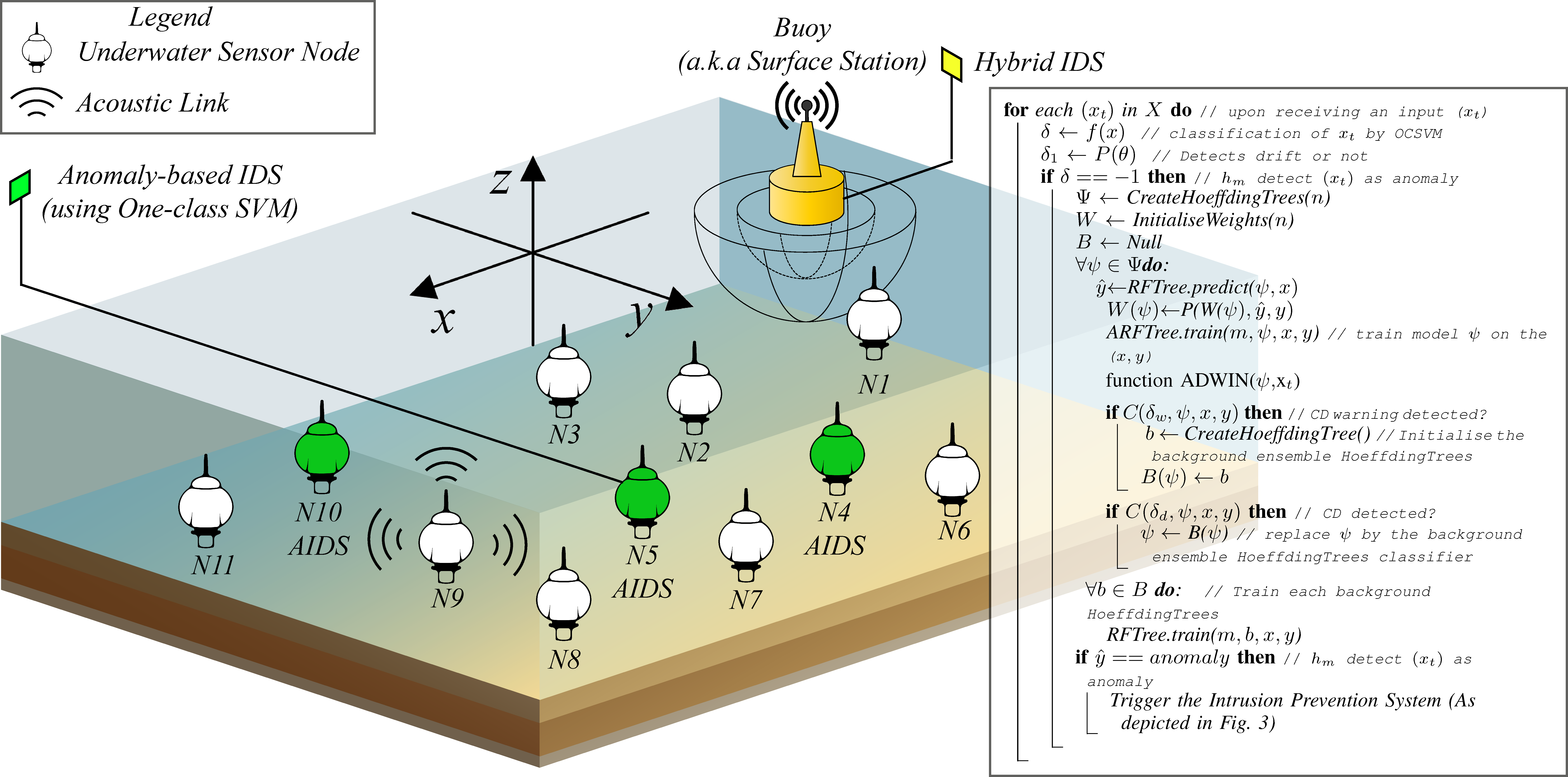}
    \caption{Decentralised architecture of the proposed solution.}
    \label{fig:UW-ASN}
\end{figure*}

UW-ASNs can vary in architecture depending on the use case, such as static two-dimensional UW-ASNs, static three-dimensional UW-ASNs, and three-dimensional networks. The mode of communication used for UW-ASNs is acoustic waves which use carrier waves to transmit data through modulation such as amplitude and frequency. Acoustic waves can cover long distances underwater (more than 100 kms) \cite{su2012acoustic}. Figure \ref{fig:UW-ASN} depicts an UW-ASN. UW-ASNs face challenges due to hardware limitations of nodes, acoustic propagation, and unstable underwater environment. UW-ASNs are deployed in constantly evolving data environments (e.g., due to sensor ageing, underwater current etc.). Sensors and actuators in UW-ASNs are resource constrained (e.g., limited energy resource, computational power, and storage capacity). The acoustic waves are influenced by high and variable path loss, Doppler spread, latency due to the propagation delay, limited bandwidth, lower data rate, noise and speed.

The existing routing protocols \cite{han2015routing} for UW-ASNs, which help to communicate and share information among the nodes, are Hop-by-Hop dynamic addressing-based (H2-DAB) \cite{ayaz2009hop}, geographic and opportunistic routing with depth adjustment-based topology control for communication recovery (GEDAR), energy-optimised path unaware layered routing protocol (E-PULRP) \cite{gopi2008energy}, oower-efficient routing (PER) and vector-based forward (VBF). H2-DAB routing protocol uses a dynamic addressing scheme among the sensor nodes to communicate and does not require localisation information. GEDAR implements a greedy, opportunistic mechanism to route data packets for communication. E-PULRP is a hybrid routing protocol that consists of layering and communication phases. In the PER protocol, a fuzzy logic inference system is used for forwarding packets towards the sink node. Hence, the forwarding tree-trimming approach is adopted to prevent the spread of forwarded packets. 

The routing protocols of UW-ASNs are exposed to various attacks related to confidentiality, integrity, and availability (CIA) of actuated and sensed information \cite{bhairavi2018secure}. The nature of these routing threats can be categorized as passive or active \cite{yang2019challenges}. The passive routing attacks include eavesdropping attacks and traffic analysis
attacks. The active attacks include denial of service (DoS) attacks, repudiation attacks and routing attacks. DoS attacks are the most dangerous and challenging to detect \cite{zlomislic2014denial}. Some existing DoS-based attacks can be classified as blackhole, grayhole, flooding, scheduling, Sybil/wormhole, and low-rate flooding attacks \cite{ahmad2021classification}. In case of a blackhole attack, the compromised sensor node, which acts as the forwarding agent, drops the collected packets, increasing packet loss significantly. A grayhole attack is a transformation of blackhole attack, where the compromised node strategically forwards or drops the packets to minimise the chance of getting exposed. A flooding attack floods the child node with packets sent from a malicious node or group of malicious nodes (Distributed Denial of Service attack) to reduce the bandwidth and exhaust energy.  

Several defence mechanisms have been developed to maintain the CIA in UW-ASNs. The system's confidentiality is achieved by implementing the Cipher Text Stealing (CTS) encryption technique \cite{rogaway2012security}. Intrusion Detection System (IDS) and Intrusion Prevention System (IPS) achieve the integrity and availability of the system. Although IDS aims to detect and identify abnormal activities, it does not mitigate detected anomalous activities. Hence, researchers have developed IPS to not only detect intrusions but also prevent the compromised nodes from taking any further actions. IDS are mainly classified based on their detection technique which can be classified based on the data source: host-based or network-based; detection technique: signature-based or anomaly-based; architecture: centralised or de-centralised; and environment: wired, wireless or ad-hoc network. Network-based IDSs gather data by analysing the network traffic, whereas host-based IDSs deploy a host in the form of agents, which runs on a monitored device. Signature-based IDSs match the pattern of the monitored network's data to a database of known attack signatures for classification. However, this approach fails when a zero-day (unseen) attack occurs. An anomaly-based IDS establishes a normal baseline set dynamically and monitors the network to compare them against the baseline for anomalous behaviour. This type of IDSs can handle unknown attacks better; however, it increases the chances of false positives or alarms. Some IDSs use machine learning/deep learning and statistical methodologies to train the model to classify and detect attacks by processing and analysing the data from a network \cite{halimaa2019machine}. Such IDSs lack adaptivity to the changes in the evolving underwater environment. In this paper, we develop a new adaptive IPS for UW-ASNs to secure them against blackhole, grayhole, and flooding attacks. We propose an adaptive IDS and IPS system suitable for the targeted UW-ASNs using a hybrid model that includes adaptive random forest (RF) and one-class support vector machine (SVM) algorithms for concept drift detection. This hybrid adaptive model outperforms the existing standard ML-based IDS and IPS solutions. 

\subsection{Desirable Properties}
\label{subsec:Desirable_Properties}

Considering the discussed challenges in UW-ASN, any defensive system should address the following Desirable Properties (DPs):

% Our proposed adaptive IDS approach aims to achieve the following Desirable Properties (DPs):

\begin{itemize}

  \item \textbf{DP1 (Zero-day intrusion detection):} The defense system is expected to detect known and previously unseen intrusions accurately.

 \item \textbf{DP2 (Adaptive):} The defense system is expected to be adaptive against the evolving underwater environment to efficiently manage the imbalanced streaming data.

 \item \textbf{DP3 (Out-Of-Distribution data detection):} The defence system should be able to detect the time and place (when and where) of shifts in data distribution (a.k.a. concept-drift) in the evolving data stream.

 \item \textbf{DP4 (Scalable):} The defence system is expected to be generalised and maintain its performance against various scaled underwater network infrastructures (when the UW-ASN is scaled up with more sensor nodes).

 \item \textbf{DP5 (On-the-fly detection):} The defence system is expected to detect threats on the fly (because detecting the threats in real-time makes the system efficient in preventing the adversary from taking further actions).

 \item \textbf{DP6 (Lightweight):} The defence system is expected to be lightweight and computationally efficient since the UW-ASN senor nodes are resource constrained.  
 
 \item \textbf{DP7 (Intrusion Prevention System):} The defence system should be integrated with a self-threat prevention system to prevent the adversary from taking further actions.
 
\end{itemize}

\subsection{Motivation and Contribution}
\label{sec:Problem Statement}
% Even though for an IDS, we expect it to support (\emph{DP1}-\emph{DP7}), however, based on our research (as shown in Section \ref{sec:Related Work}), none of the existing IDS can satisfy all the desirable properties mentioned in Section \ref{subsec:Desirable_Properties}.
% }
To the best of our knowledge (as shown in Section \ref{sec:Related Work}), existing IDSs in the literature cannot ensure all of the above-mentioned desirable properties (\emph{DP1}-\emph{DP7}) for UW-ASNs. 
Moreover, the occurrence of an intrusion is, in general, a rare incident (with respect to the volume of normal observations over the entire monitoring period) which makes the streaming data imbalanced and skewed toward the majority class. Hence, such an imbalanced streaming data environment causes an additional challenge for any learning and monitoring agents.

%\subsection{\textcolor{blue}{Contributions}}
\label{sec:Contribution}

In order to mitigate the existing challenges discussed in Section \ref{sec:Problem Statement}, an incremental security system is required to adapt to changes in data distribution (a.k.a. concept drifts) on-the-fly. Since it is not feasible for any security system to obtain and accommodate the entire normal and malicious activities, the development of an incremental and generalised security system is required to accurately classify out-of-distribution (OOD) data. In this context, due to the lack of adaptivity in the existing IDS for the evolving environment of UW-ASNs, we propose a robust and adaptive IPS to protect UW-ASNs against blackhole, grayhole and flooding attacks. The proposed IPS aims to achieve all desirable properties.

This paper makes the following contributions:
\begin{itemize}
  \item A new robust hybrid incremental IDS to detect UW-ASN routing attacks. The proposed scheme can detect and adapt to shifts in data streams (a.k.a. concept drifts) on-the-fly to maintain its detection performance. 

  \item The \emph{first} incremental cryptography-based IPS, which is lightweight and isolated against an external adversary. The proposed IPS can avoid the negative impacts of false positives.

  \item The \emph{first} solution to identify and mitigate grayhole, flooding, and blackhole routing attacks in UW-ASN environments.
  
  \item A generated dataset\footnote[2]{\textbf{Our datasets and source codes are available in the link below: \href{https://drive.google.com/drive/folders/11d6tZAOkqvdrj57A0OFAlgf7YlQgzTzz}{drive.google.com/drive/folders/11d6tZAOkqvdrj57A0OFAlgf7YlQgzTzz}}} with 16$\sim$64 nodes for UW-ASNs, for the research community to use as a benchmark (UW-ASN dataset).
  
  \item Benchmarked the performance of the proposed scheme against most of the state-of-the-art machine learning classifiers.
 
\end{itemize}

\subsection{Organisation}
\label{sec:Organisation}
\begin{figure}[t]
    \centering
    \includegraphics[width=9cm]{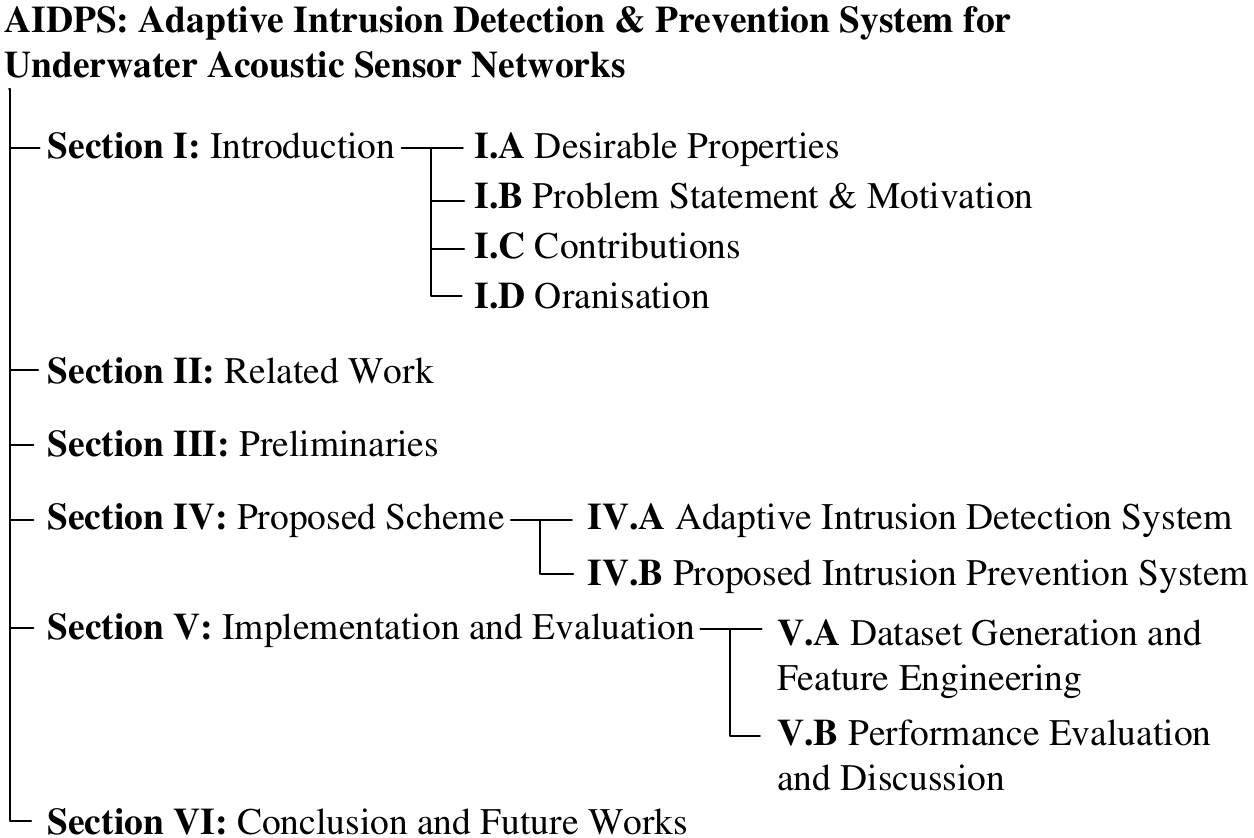}
    \caption{Organization of the paper.}
    \label{fig:Structure}
    % \vspace{-3mm} %Put here to reduce too much white space after your table 

\end{figure}

The rest of the paper is organised as follows. In Section \ref{sec:Related Work}, we discuss the related works. In Section \ref{sec:Preliminaries}, we present the preliminaries. Section \ref{sec:Proposed_Work} presents the proposed scheme. Section \ref{sec:Implementation and Evaluation} describes our implementation and evaluation details. Section \ref{sec:Conclusion} concludes the paper and lists possible future work in this area. The organization of the paper is illustrated in Fig. \ref{fig:Structure}.

\section{Related Work}
\label{sec:Related Work}
The lack of an efficient IDS, along with the external challenges (such as unstable underwater environment and UW-ASN sensor resource constraints) posed by the unstable nature of the underwater environment, makes UW-ASNs prone to vulnerabilities. Various IDSs have been proposed to secure UW-ASNs. However, most fail to achieve all the desirable properties required to build an efficient IDS for UW-ASNs. Table \ref{tab1} compares related works to our approach with respect to the desirable properties.

%The challenges mentioned above have encouraged researchers to develop security mechanisms to address the vulnerabilities related to the wireless network. 
The use of mobile agents to detect sinkhole attacks in wireless sensor networks (WSNs) is proposed in \cite{hamedheidari2013novel}. The proposed mechanism uses mobile agents to make sensor nodes aware of their trusted neighbours by not listening to traffic coming from malicious sensor nodes. Instead of the traditional client-server-based processing, it leverages the use of mobile agents and traverses through the network either periodically or on demand, bringing the functionality to the data rather than the other way around. The proposed mechanism increases scalability and keeps track of sensor network resource constraints, making them lightweight and energy efficient. 
This scheme can ensure the scalability (\emph{DP4}) property and is resource efficient, but it is limited to just sinkhole attacks and built for WSNs. Also, this scheme fails to cover all the desirable properties required by an efficient IDS. It fails to prevent zero-day intrusions (\emph{DP1}) and increases the detection time of the malicious node (\emph{DP5}), as the agent needs to reach the sensor nodes to detect intrusion.

To overcome the properties lacked by \cite{hamedheidari2013novel}, researchers in  \cite{ahmad2022comprehensive} leverage deep hybrid learning (DL) models across benchmarked datasets and analyse their performance. Individual DL classifiers implemented were Multi-layer Perceptron, CNN, and LSTM. Hybrid classifiers implemented were  Autoencoder and Temporal Convolutional Network, Autoencoder and LSTM, Autoencoder and Bi-Directional Recurrent Neural Network, Autoencoder and Bidirectional LSTM, and CNN and LSTM. 
Even though the DL-based security mechanism is adaptive (\emph{DP2}), the models cannot detect intrusions on-the-fly (\emph{DP5}), as the DL-based models require a good amount of training time. The model needs to be trained on the entire dataset every time there is a change in the data distribution, thereby increasing the training time. In addition, the models are not lightweight (\emph{DP6}). Also, the datasets considered are specific to IoT networks on which the models have been trained and evaluated. The datasets considered are also aged and thus do not consider recent attacks. Since the model does not implement out-of-distribution data detection (\emph{DP3}), it would require continuous monitoring and re-training of the model whenever the underlying data distribution changes, thereby increasing the cost of implementing such models in an evolving data stream.

To handle the problem of Out-Of-Distribution data detection (\emph{DP3}), a network-based IDS (NIDS) has been proposed in \cite{ramkumar2022intrusion} using Spark's master-slave node architecture. The data nodes containing the data perform feature selection in different slave nodes using RV coefficient-based hybrid feature fusion, which is designed by incorporating the wrapper, class-wise information gain, and Canberra distance. The unique features selected then undergo a process of Data Augmentation using oversampling on the slave node. The intrusion detection classification and training are done in the master node, which uses the DRN classifier.
This approach has the potential to handle out-of-distribution data detection (\emph{DP3}) and is specific for Internet applications. The proposed security mechanism can achieve on-the-fly detection (\emph{DP5}) of intrusions. However, the datasets used are not specific to UW-ASNs, and no intrusion prevention mechanism has been proposed. 

Another approach for designing an efficient IDS combines packet-based and flow-based intrusion detection techniques, which makes the IDS hybrid by considering both the traffic flow and packet analysis. The authors in \cite{qiu2022hybrid} propose an IDS that uses Dempster-Shafer theory (DST-IDS).  DST-IDS is an ensemble method that takes traffic flow information and the first $N$ packets as input. Both the traffic flow predictions and packet-based IDS are fused to get the final detection result. A data collection and processing tool was proposed to reduce the processing time for massive data volumes. In addition, it was designed to work with heterogeneous data distributions to provide scalability to the DST-IDS.
Though this technique stands out well regarding scalability (\emph{DP4}) and on-the-fly detection (\emph{DP5}), it fails to achieve other desirable properties.

In summary, the papers mentioned above introduce different techniques for developing an effective IDS. However, each system lacks some desirable properties, as depicted in Table \ref{tab:Related Works}. Also, these IDSs are explicitly built for wireless terrestrial networks. They do not consider the challenges of UW-ASNs and the acoustic mode of communication, making them unsuitable for UW-ASN environments. To the best of our knowledge (as shown in Table \ref{tab:Related Works}), there is no IDS with all the above-mentioned desirable properties (\emph{DP1}-\emph{DP7}) available for UW-ASNs. We believe that our proposed scheme will contribute to fill this gap and be used by the research community for future works related to UW-ASNs.

% To the best of our knowledge (as shown in Table \ref{tab:Related Works}), existing IDS's in the literature cannot ensure all of the above-mentioned desired properties (\emph{DP1}-\emph{DP7}) for UW-ASNs

%%%%%%%Related Works Table%%%%
\begin{table*}[ht!]
\caption{Related Works}
\centering

%|p{0.2\linewidth}|M{0.11\linewidth}

\begin{center}

\scalebox{.8}{\begin{tabular}{|c|p{0.6\linewidth}|p{0.15\linewidth}|c|c|c|c|c|c|c|c|}
\hline
\textbf{Scheme}&\textbf{Approach}&\textbf{Threat}&\multicolumn{7}{|c|}{\textbf{Desirable Imperative Features}} \\
\cline{4-10} 
\textbf{} & \textbf{} & \textbf{} & \textbf{\textit{DF1}}& \textbf{\textit{DF2}}& \textbf{\textit{DF3}} & \textbf{\textit{DF4}}
& \textbf{\textit{DF5}}& \textbf{\textit{DF6}} & \textbf{\textit{DF7}}
\\
\hline
\cite{hamedheidari2013novel} & Novel agent-based approach to detect sinkhole attacks in wireless sensor networks (WSNs) & SA
& $\times$ & $\times$ & $\times$ & $\checkmark$ & $\times$ & $\checkmark$ & $\checkmark$ \\
\hline

\cite{ahmad2022comprehensive} & Deep learning benchmark for IoT IDS & D/N
 & $\times$ & $\checkmark$ & $\times$ & $\checkmark$ & $\times$ & $\times$ & $\times$ \\
\hline

\cite{ramkumar2022intrusion} & RV coefficient+Exponential Sea Lion Optimisation-enabled Deep Residual Network (ExpSLO-enabled DRN) based on spark architecture & D/N
& $\times$ & $\times$ & $\times$ & $\checkmark$ & $\checkmark$ & $\times$ & $\times$\\
\hline
    
\cite{qiu2022hybrid} & Hybrid IDS based on Dempster-Shafer evidence
theory & DDoS
& $\times$ & $\times$ & $\times$ & $\checkmark$ & $\checkmark$ & $\times$ & $\times$\\
\hline

Proposed & Hybrid IDS using One Class SVM and Incremental adaptive random forest using ADWIN and & BA, GA and FA & $\checkmark$ & $\checkmark$ & $\checkmark$ & $\checkmark$ & $\checkmark$ & $\checkmark$ & $\checkmark$\\
Scheme &  kdqTree concept drift detectors &   &  &  &  &  &  & & \\
\hline
% \vspace{1mm}

\multicolumn{10}{l}{$^{\mathrm{*}}$\textbf{D/N:} Different Network-technology. \textbf{DP1:} Zero-day intrusion detection. \textbf{DP2:} Adaptive. \textbf{DP3:} Out-Of-Distribution data detection. \textbf{DP4:} Scalable. \textbf{DP5:} On-the-fly detection.\textbf{ DP6:} Lightweight. 
}
\\
\multicolumn{10}{l}{$^{\mathrm{}}$\textbf{ DP7:} Intrusion Prevention System. \textbf{ BA:} Blackhole Attack. \textbf{GA:} Grayhole Attack. \textbf{FA:} Flooding Attack. \textbf{SA:} Sinkhole Attack. 
\textbf{DDoS:} Distributed Denial of Service Attack.
}
% \\
% \multicolumn{10}{l}{$^{\mathrm{}}$

% }

\end{tabular}}

\label{tab1}
\end{center}
\label{tab:Related Works}%
\end{table*}

%\subsection{Motivation and Contribution}

\section{{Preliminaries}}
\label{sec:Preliminaries}

%\IEEEpubidadjcol
This section introduces the background concepts relevant to the paper. To begin with, we provide an introduction to the various routing protocols employed in UW-ASN. Specifically, we focus on the Vector Based Forward protocol, which we utilized in our experimental setup. Subsequently, we outline the diverse attacks conducted against UW-ASN. Then, we present a comprehensive overview of different types of IDS and highlight their significance in safeguarding UW-ASNs. Additionally, we introduce the concept of incremental machine learning, which serves as a fundamental component of our proposed system. Lastly, we introduce various techniques for detecting concept drift and underscore their importance in the context of our research.

\subsubsection{\textbf{Routing Protocols}} 
\begin{figure}[htp]
    \centering
    \includegraphics[width=86mm]{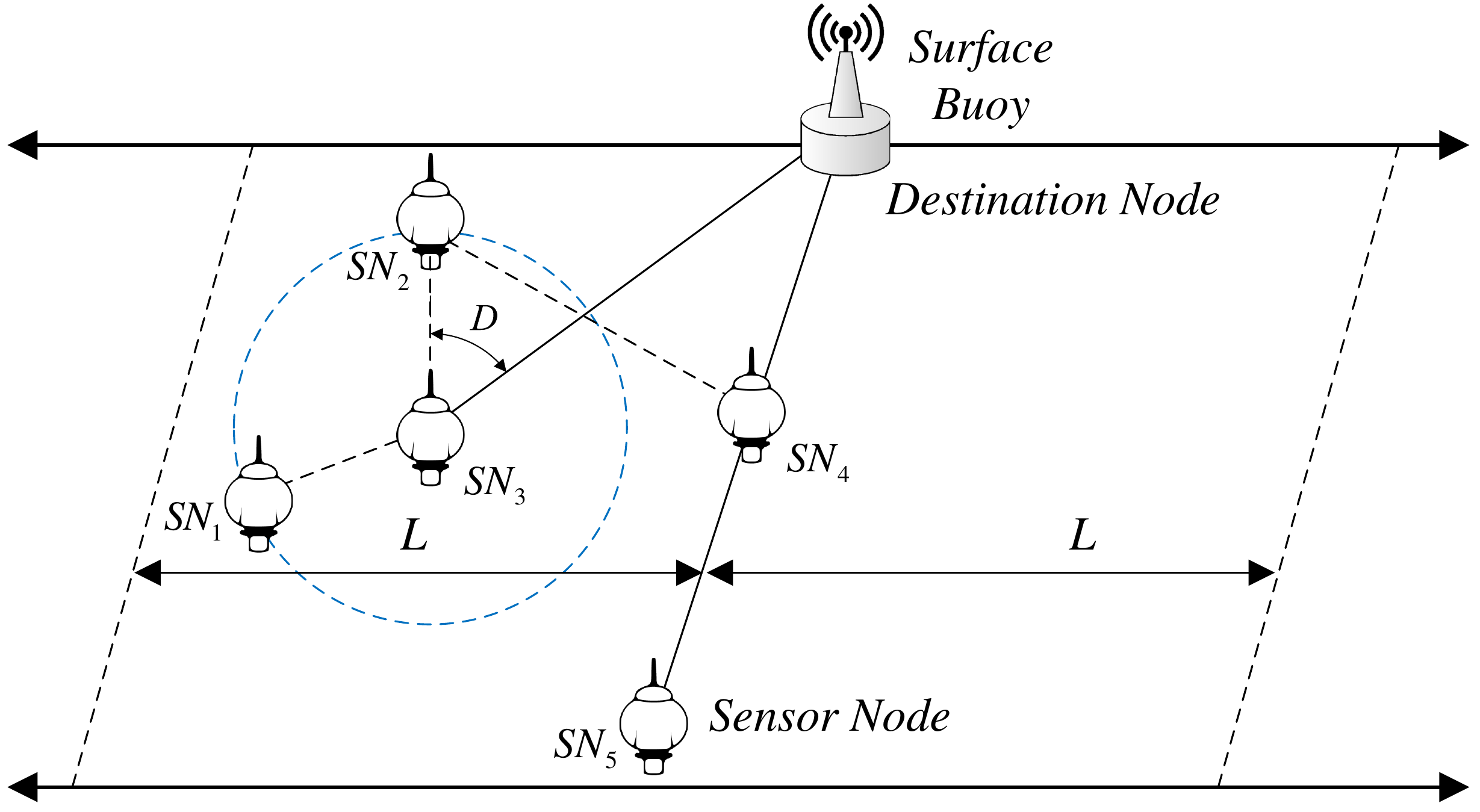}
    \caption{Vector-based forward (VBF) routing.}
    \label{fig:VBF}
\end{figure}

A routing protocol selects a suitable route or path for the data to travel from its source sensor node to its destination sensor node. In an UW-ASN environment, the data must be sent from the sensor node to the surface node. The surface node sends the data to the surface station or base station. This connects the underwater sensor nodes to other networks. To achieve reliable communication, the design of a routing protocol becomes critical. Among the existing routing protocols for UW-ASN (e.g., H2-DAB, E-PULRP, GEDAR, VBF and PER) \cite{han2015routing}, VBF is an efficient routing protocol in underwater environments as it considers the energy constraints and node mobility issues. VBF is a position-based routing approach \cite{xie2006vbf}, ensuring a robust, scalable and energy-efficient routing, and addressing the node mobility issue. Only the nodes close to the sensor node that is generating the packet to be sent to the next node (a.k.a. vector) will forward the packets (as shown in Fig. \ref{fig:VBF}). Therefore, only a tiny fraction of the nodes communicate, preserving their energy resources and reducing network overhead. The VBF protocol also implements a self-adaptation algorithm to adjust the forwarding policy based on local information, forcing the algorithm to consider the density of the neighbourhood nodes to modify its forwarding policy for energy efficiency. This paper uses VBF as the routing protocol \cite{xie2006vbf}.

 \subsubsection{\textbf{Attacks against UW-ASNs}}
 UW-ASNs have many applications, such as marine ecosystem monitoring, international water border surveillance, underwater equipment monitoring, and natural calamity detection. They are often deployed in unprotected and hostile environments, which makes them vulnerable to attacks. DoS attacks are a common and dangerous attack for UW-ASNs which interrupt the service and make its functionality void to cause a negative impact on the network's availability. UW-ASNs inherit various types of routing attacks from Wireless Sensor Networks (WSNs), namely blackhole attacks and grayhole attacks. Furthermore, the adversary can generate various flooding attacks (e.g. Low-Rate DoS and Distributed DoS) in this network
 \cite{ahmad2021classification}. Attacks that have been considered as part of our experiments are:
\begin{itemize}
 \item Blackhole attack: A blackhole attack occurs when an intermediary re-programs a sensor node or a set of sensor nodes to drop the packets instead of forwarding them to its neighbouring node \cite{ahmad2021classification}.

 \item Grayhole attack: Grayhole attacks are a type of DoS attack which implements selective forwarding. The compromised sensor node selectively drops some packets and forwards the remaining packets to the destination nodes \cite{ahmad2021classification}.

 \item Flooding attack: Flooding, as the name suggests, is a type of DoS attack which targets a sensor node and increases the traffic in that node. The high traffic volume can be sent by a single malicious node or a group of nodes to decrease the overall performance of the UW-ASN \cite{ahmad2021classification}.

\end{itemize}
To generate the dataset, these attacks were implemented by changing the vector-based-forward routing protocol provided by the underwater-sensor patch in Network Simulator 2 (NS2). 16 sensor nodes were considered as part of the UW-ASN for generating the dataset and 64 sensor nodes for generating the out-of-distribution dataset, to test {DP3} of the desirable property.

\subsubsection{\textbf{Intrusion Detection System}} An IDS is used to detect intrusions or threats. An IDS can either be Host-Based IDS or NIDS.  
 
 \begin{itemize}
 \item Host-Based IDS: A host-based IDS protects a particular sensor node against internal and external threats. This type of IDS is capable of monitoring network traffic to and from the sensor node. It gets insight into the host's in-state. However, its visibility is limited to only the host sensor node.

 \item Network-Based IDS: A NIDS solution is designed to monitor an entire network, giving it visibility into all traffic flowing through it. This broader viewpoint provides the ability to detect threats in a wider area. However, they lack visibility inside the sensor nodes.
 
 \end{itemize}

IDSs' detection techniques can also be classified as signature-based detection, anomaly-based detection, and hybrid detection. 
 
 \begin{itemize}
 \item Signature-based detection: Signature-based IDS (SIDS) recognise threats with the help of signatures of known threats. This technique reduces false positives but is vulnerable to zero-day vulnerabilities.

 \item Anomaly-based detection: Anomaly-based IDS (AIDS) build a bounding wall of normal behaviour. All future behaviours are compared to this 'normal' behaviour. This technique helps detect zero-day vulnerabilities but increases false positives.

 \item Hybrid detection: A Hybrid IDS (HIDS) uses a combination of both signature-based and anomaly-based detection techniques. This helps the HIDS to reduce false positives and detect zero-day vulnerabilities.
 
 \end{itemize}

 \subsubsection{\textbf{Incremental Machine Learning}}
 \label{sec:}
 Due to the evolving data environment of UW-ASN, the dataset does not remain static. Such a dataset involves data streams that can change over time. Having a model trained over such a dataset will yield poor results whenever the characteristics of the dataset change. Incremental machine learning can continuously learn from the changing stream of data and can maintain previously learned knowledge. Real-world applications use it to learn how data arrives over time in ever-changing environments \cite{syavasyareview}. This makes our proposed IDS accurately identify known and previously unseen intrusions (DP2). Incremental machine learning techniques used in our experiment are:
\begin{itemize}
 \item \textbf{Adaptive Random Forest (ARF) Classifier}: Adaptive Random Forest Classifier is an ensemble algorithm which implements a group of Hoeffding trees. The final classification is computed by taking the votes from all the Hoeffding trees, where the class with the most votes becomes the final classification result. To handle drifts in the evolving data environment, a concept drift detection algorithm is coupled with the adaptive ensemble algorithm \cite{gomes2017adaptive}. We connected the ADWIN concept drift detection algorithm with ARF for our experiments. The concept drift detection algorithm provides the algorithm with a `warning' signal when the drift is initially detected and a `flag' signal when the drift becomes significant. As soon as a `warning' is detected, the ARF trains a set of background Hoeffding trees, replaces the foreground forest when the signal changes to `flag', and stores the existing forest to be used in case the current scenario in the data environment reappears. ARF induces diversity through re-sampling and randomly selecting subsets of features for node splits, which effectively helps the algorithm to handle class imbalance. 
 
 \item \textbf{Hoeffding Adaptive Tree (HAT) Classifier}: HAT is an incremental decision tree that uses ADWIN concept drift detection to monitor the performance of branches on the tree. HAT adaptively learns from data streams that change over time and replaces the branches with new branches when their accuracy decreases \cite{bifet2009adaptive}.

\end{itemize}

\subsubsection{\textbf{Concept Drift}}
\label{sec:preliminaries_concept_drift}
Due to the evolving data environment of UW-ASN, the properties of the dependent variables change over time. The model built using these dependent variables will decay in accuracy if the change becomes significant. The changes can be classified as Sudden/Abrupt, Incremental/Gradual or Recurring/Seasonal. An example of concept drift is when the underlying data distribution changes over time due to an influence. If the data engineering process is not strictly static, the changing underlying patterns in the data can be captured over time \cite{gama2014survey}. To be able to handle the concept drift effectively (DP3), different drift detection algorithms can be employed \cite{lu2018learning} \footnote[3]{\textbf{The other concept drift detection algorithms are discussed in Appendix D of the Supplementary Material.}}:

\begin{itemize}
 \item \textbf{Adaptive Windowing (ADWIN) \cite{ikonomovska2011adaptive}}: ADWIN is based on calculating a window size ($W$) which grows dynamically unless the data pattern changes and shrinks when a change is detected. Based on the distribution of the data, the algorithm then attempts to find two subwindows of $W$ ($w_{0}$ and $w_{1}$), which have different averages.   
 \item \textbf{Drift Detection Method (DDM) \cite{costa2018drift}}: This method is based on the Probably Approximately Correct (PAC) learning model premise that a learner's error rates decrease as more samples are analysed, as long as the data distribution remains stationary. Changes are detected if the algorithm detects an increase in error rate that exceeds a calculated threshold.

 \end{itemize} 

% \textcolor{blue}{The other concept drift detection algorithms are discussed in Appendix D of the Supplementary Material.} \\

\section{Proposed Scheme}
\label{sec:Proposed_Work}

The proposed scheme employs a hybrid adaptive IDS (Section \ref{sec:ids}) and a cryptographically secured IPS (Section \ref{Crypto-IPS}) for UW-ASN. Together, this forms the Adaptive Intrusion Detection and Prevention System (AIDPS).
Algorithm \ref{Alg:hybrid-ids} shows our proposed scheme.

\begin{figure*}[htp]
    \centering
    \includegraphics[width=15cm]{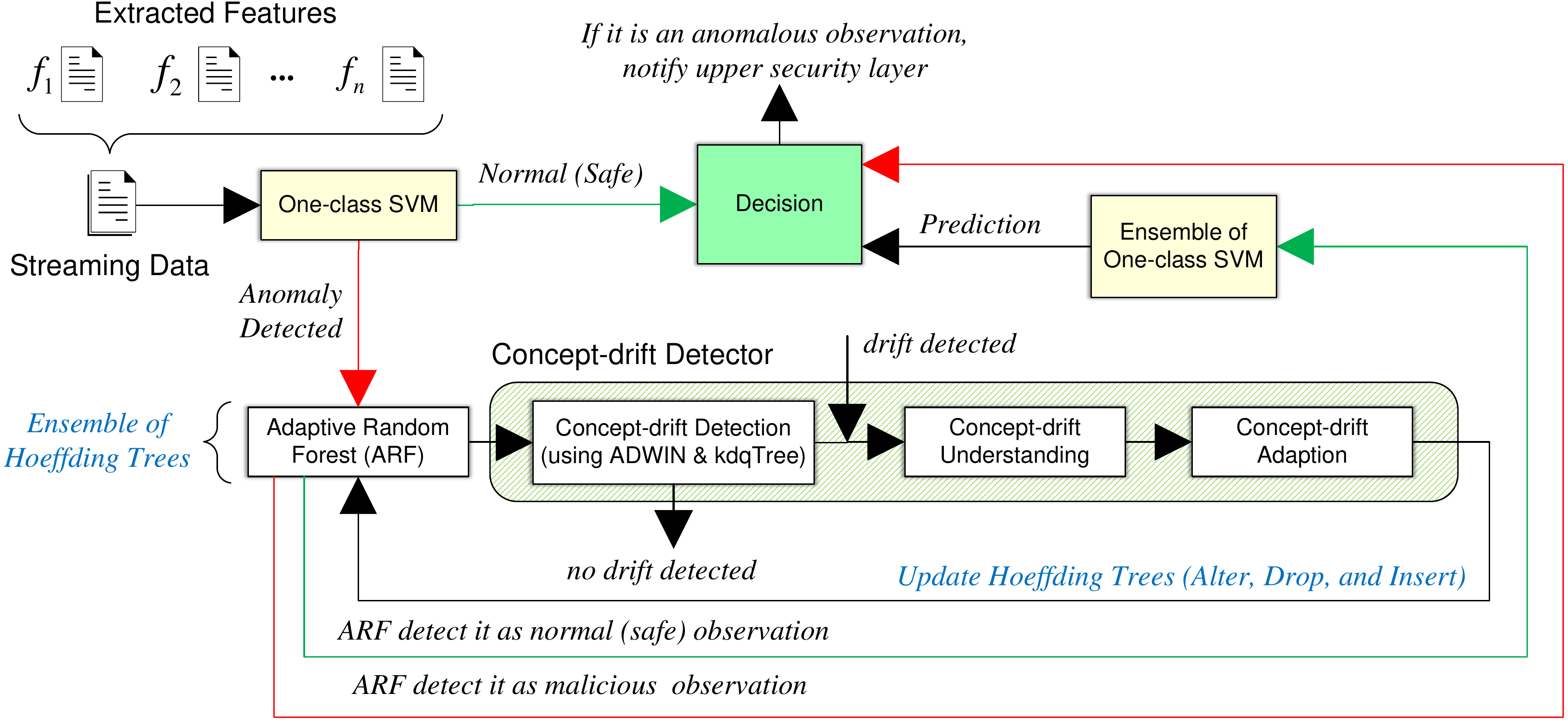}
    \caption{Adaptive intrusion detection system: proposed scheme.}
    \label{fig:Adaptive Intrusion Detection System (IDS): Proposed Scheme}
\end{figure*}

\subsection{Adaptive Intrusion Detection System}
\label{sec:ids}

The proposed IDS is a combination of anomaly-based IDS and signature-based IDS which makes it hybrid.

\subsubsection{Anomaly-based IDS}
\label{sec:Anomaly-based IDS}
\begin{figure}[t]
    \centering
    \includegraphics[width=8.4cm]{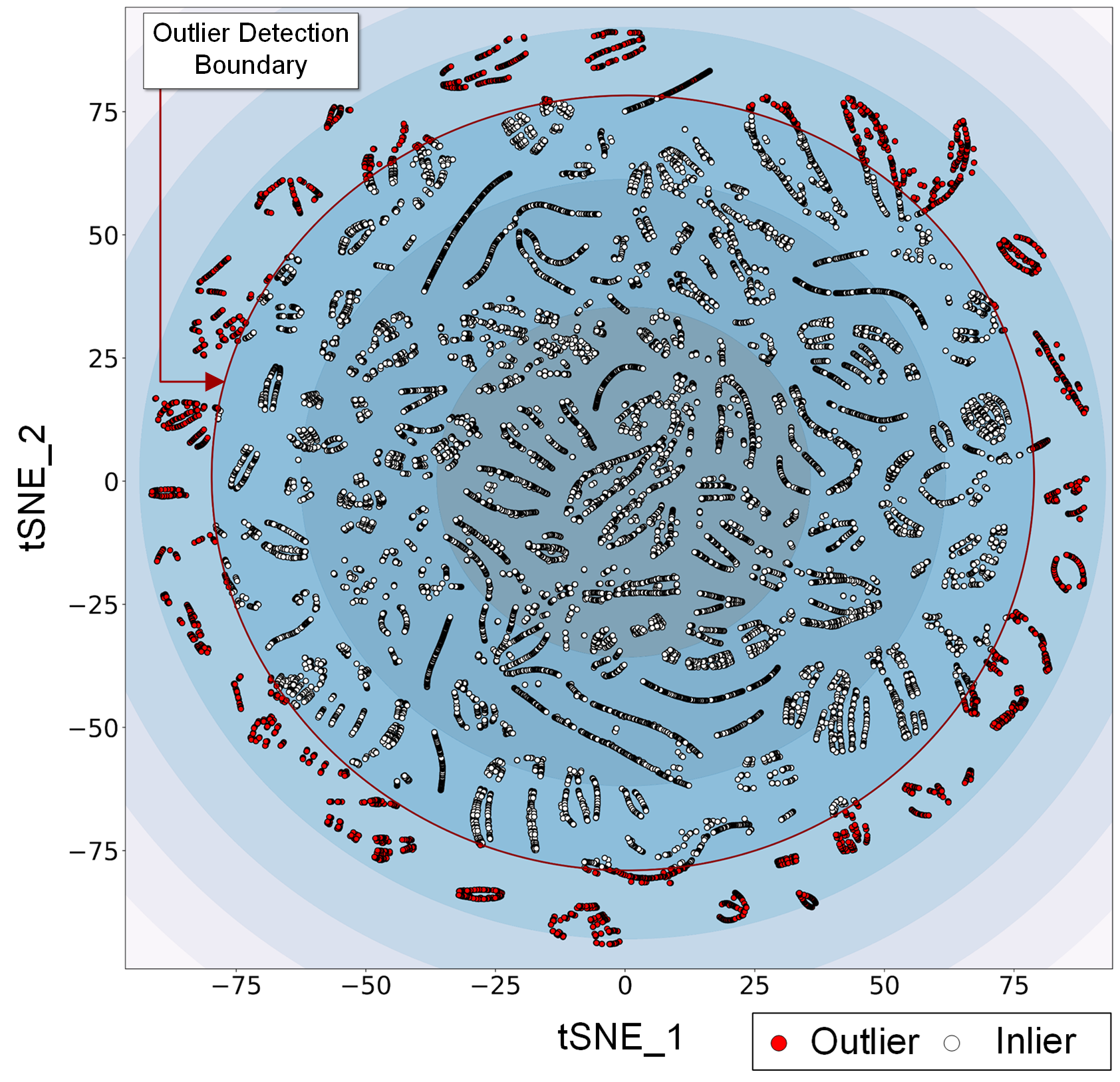}
    \caption{One-class support vector machine (OCSVM) on a t-SNE projection ($\nu$ and $\gamma$ assigned to 0.01 and 0.3 respectively).}
    \label{fig:One-Class SVM (OCSVM)}
\end{figure}

Due to their resource-constrained nature, underwater nodes are unable to accommodate computationally complex algorithms. Hence, in the proposed scheme, we develop an anomaly-based IDS monitoring agent to detect anomalous activities. One-class support vector machine (OCSVM) has been used to detect abnormal behaviour in the incoming data. OCSVM learns a decision function using its semi-supervised algorithm and classifies new data as similar or different to the training set \cite{wang2004anomaly}. Anomaly-based IDS is used to classify the new data stream as either normal if it lies within the decision boundary or anomalous if it lies outside the decision boundary. Figure \ref{fig:One-Class SVM (OCSVM)} shows the OCSVM classification of normal and abnormal behaviour on the data points. The dataset was projected onto two dimensions using t-distributed Stochastic Neighbor Embedding (t-SNE) dimensionality reduction \cite{hu2022t}.  
The parameters used by OCSVM are \textit{$\nu$}, \textit{$\gamma$} and \textit{kernel}. The parameter $\nu$ is used to specify the percentage of anomalies. The kernel is used to identify the kernel type and also maps the data to a higher dimensional space for the SVM to draw a decision boundary. The parameter $\gamma$ is used to set the kernel coefficient.
For our experiment, the decision boundary for OCSVM has been set to $\nu = 0.01$ and $\gamma = 0.3$. The outcome of OCSVM is bipolar, where -1 are the outliers (shown in red) and +1 are the inliers (shown in white) predicted using the OCSVM's decision boundary. Section \ref{sec:Performance_Evaluation} (Experiment 1) discusses the evaluation of the OCSVM model on the UW-ASN dataset that we have generated in this article. The details of the UW-ASN dataset generation and feature engineering has been discussed in Section \ref{sec:Dataset Generation and Feature Engineering}.

\subsubsection{Hybrid IDS}
\label{sec:Proposed Incremetanl IDS}
To accurately detect previously known attacks, it is essential to have a signature-based IDS. The signature-based IDS, along with the concept drift detection algorithm, can handle the incoming streaming underwater data and accurately find known signatures of attacks. When coupled with an anomaly-based IDS, this also helps to detect unknown or previously unseen attacks. The entire system works in tandem as a hybrid IDS, which can achieve the desirable properties of DP1, DP2 and DP3. 

The anomalous data points from the first (OCSVM-based) anomaly-based IDS were sent to the ARF classifier algorithm, an ensemble of Hoeffding trees (HT). ARF is an incremental machine learning algorithm widely used for evolving data streams. ARF, in turn, uses ADWIN and kdqTree concept drift detection algorithms, which implement error-rate-based and unsupervised detection techniques. The algorithm is implemented with the help of Evaluate Prequential technique, where ARF is trained over a subset of the data and then predicts when the ADWIN drift detector detects no drift. ARF also uses ADWIN for drift detection warning signal, and if a drift warning is seen, a new ARF (ARF1) is trained (which includes an ensemble of new Hoeffding trees) in the background. ARF1 replaces ARF as soon as ADWIN provides a drift signal, and predictions are taken from ARF1. This makes the proposed IDS adaptive (DP2). Section \ref{sec:Performance_Evaluation} (Experiment 2) discusses the evaluation of the ARF model on the UW-ASN dataset.

Figure \ref{fig:Adaptive Intrusion Detection System (IDS): Proposed Scheme}, shows the architecture of our Adaptive IDS. The OCSVM anomaly detector takes the incoming data stream as input and gives its predictions. The data stream showing normal behaviour is allowed to pass and considered part of the final prediction. Data streams showing anomaly behaviour are then passed through an ARF classifier, which implements an ensemble of Hoeffding trees. ARF implements a drift detector algorithm in our proposed scheme: ADWIN and kdqTree. If no drift is detected, the ARF algorithm's predictions are sent to OCSVM for the final prediction. However, when ADWIN detects a drift warning signal, it updates the ARF by updating the ensemble of trees used for ARF. During this stage, the predictions from the older ARF algorithm are considered part of the final prediction by ARF. When ADWIN detects a clear drift signal, it replaces the old ARF algorithm with the new updated ARF to evaluate the predictions as part of the final prediction by the ARF. The normal instances are further sent to the second anomaly-based detector, which implements an ensemble of OCSVM (Bagging of OCSVM) to check for abnormality \cite{shieh2009ensembles}. The estimated boundary is sensitive in practice as it needs to uncover zero-day (unseen) attacks. To achieve this, we use an ensemble of eleven OCSVM anomaly detectors, and the bagging concept is used to decide whether the instance is an outlier. The advantage of using the bagging OCSVM is to tighten the decision boundary and reduce false positives. 
The final decision algorithm implements a simple logic to compute the final predictions. It takes the inputs from the initial OCSVM, which shows normal behaviour, and predictions from the signature-based adaptive random forest classifier and the final ensemble of OCSVMs to give out the final predictions as either normal or attack. If the final predictions classify an instance of the incoming data stream as expected, no further actions are taken. However, if the final prediction of the incoming data stream is classified as an attack, it notifies the cryptographically secured IPS to take further action. Section \ref{sec:Performance_Evaluation} (Experiment 4) discusses the evaluation of the proposed hybrid solution on the UW-ASN dataset.

\begin{figure*}[htp]
    \centering
    \includegraphics[width=15cm]{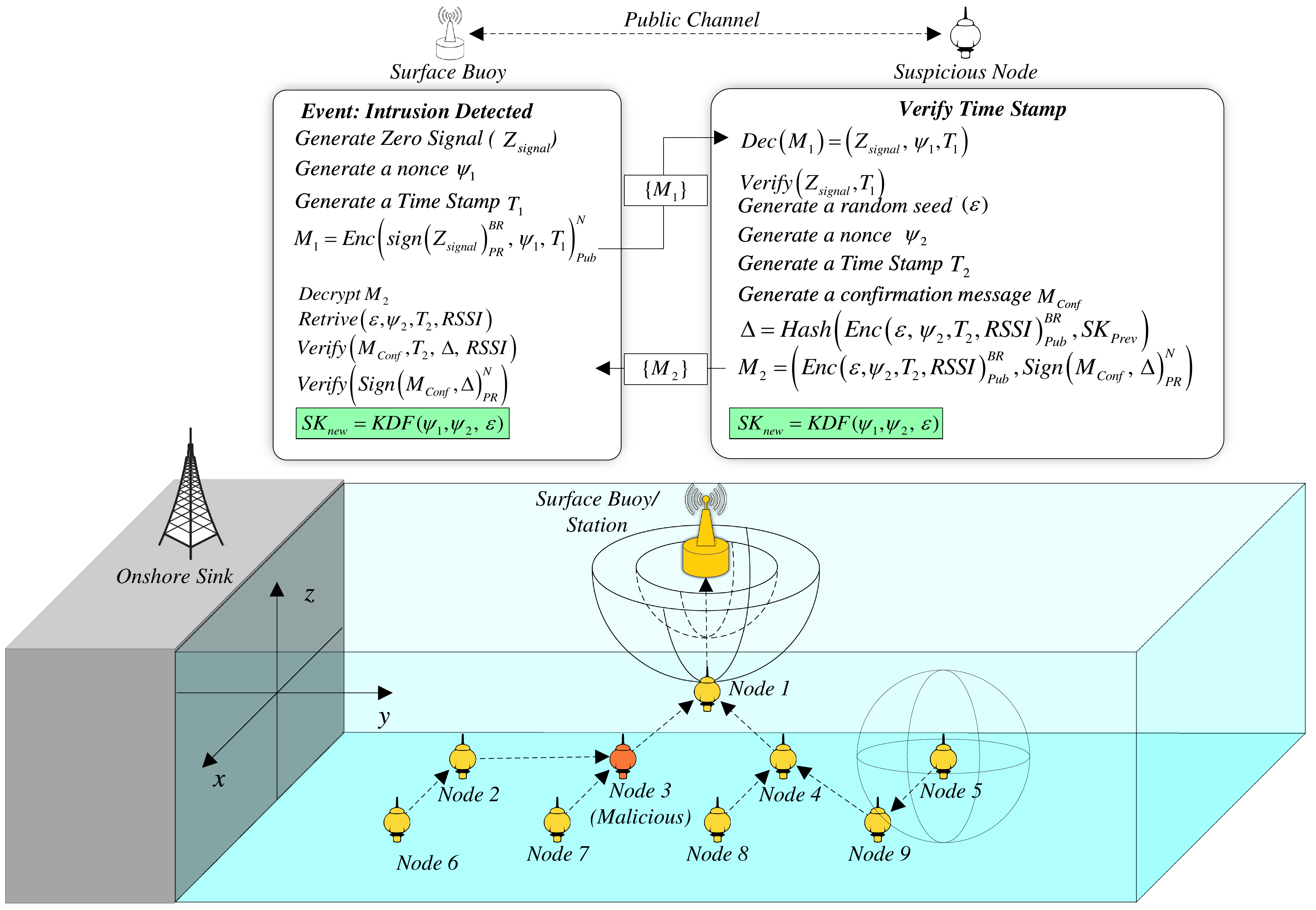}

    \caption{Intrusion prevention system.}
    \label{fig:UW-ASN-IPS}
\end{figure*}

\subsection{Proposed Intrusion Prevention System}
\label{Crypto-IPS}
In the proposed system, after the detection of anomalous activity by central IDS on the surface buoy (discussed in Section \ref{sec:ids}), the proposed intrusion prevention mechanism will be triggered. In this context, the surface buoy (border router) finds its distance from the suspicious nodes using the Received Signal Strength Indicator (RSSI). 
Upon detection of anomalous activity, the CIDS shares a new key with the suspicious node with the RSSI within the registered range. Hence, the malicious node cannot impersonate as a legitimate node (e.g., by conducting clone ID or Sybil attack) and will be isolated. Next, our proposed IPS (placed on the surface buoy) isolates potential malicious nodes and then establishes a new session key ($SK_{new}$) by replacing an old session key ($SK_{old}$) between a node and surface buoy as shown in Fig. \ref{fig:UW-ASN-IPS}. 
For the establishment of the session key, we consider that each node has a pair of keys generated through the Elliptic Curve Diffie-Hellman Key Exchange (ECDHE). Consider that the surface buoy has \{$K_{Pub}^{BR}$, $K_{Pr}^{BR}$, $K_{Pub}^{N}$\} and a node has 
\{$K_{Pub}^{BR}$, $K_{Pr}^{N}$, $K_{Pub}^{N}$\}, where $K_{Pub}^{X}$ and  $K_{Pr}^{X}$ represents a public key and private key of entity \textit{X}. As shown in Fig. \ref{fig:UW-ASN-IPS}, the surface buoy uses \textit{zero-signal ($Z_{signal}$)} to inform the node that intrusion is detected and there is a need to reset the key. In this regard, the surface buoy generates a nonce $\Psi_{1}$, timestamp T1 and computes $M_{1}=Enc(sign(Z_{signal})^{BR}_{PR}$, $\Psi_{1}, T1)^{N}_{Pub}$. Finally, the surface buoy constructs a message \{$M_{1}$\} and sends it to the node.
Upon receiving the message $\{M_{1}\}$, the node first decrypts $M_{1}$, checks the timestamp $T1 - T1^* \stackrel{?}{=} \Delta$T, and then  verifies the zero signal ($Z_{signal}$). Next, the node generates a random seed ($\varepsilon$), a random number \{$\Psi_{2}$\}, timestamp $T2$, a confirmation message $M_{Conf}$, and computes $\Delta=Hash(Enc(\varepsilon,\Psi_{2}, T_{2}, RSSI)^{BR}_{Pub},SK_{Old})$, and $Sign(M_{Conf}, \Delta)^{N}_{PR}$. Next the node constructs a message $M_{2}$ = $(Enc(\varepsilon,\Psi_{2}, T_{2}, RSSI)^{BR}_{Pub}, Sign(M_{Conf}, \Delta)^{N}_{PR})$ and sends it to the surface buoy. After receiving the message \{$M_{2}$\}, the buoy decrypts the message $M_{2}$, verifies the timestamp $T2 - T2^* \stackrel{?}{=} \Delta$T and obtains $(\varepsilon, \Psi_{2}, RSSI)$. Next the surface buoy verifies the decrypted parameters such as RSSI, $M_{Conf}$, and then verifies the signature. If the verification is successful then both the surface buoy and the node compute the new secret key $SK_{new}=KDF(\Psi_{1},\Psi_{2}, \varepsilon)$. Upon successful key generation of $SK_{new}$, both the node and the surface buoy again start secure communication with each other. This helps to achieve DP7 of the desired properties listed in Section 
\ref{subsec:Desirable_Properties}.

\begin{algorithm}[ht!]
\DontPrintSemicolon

% \small
\footnotesize
% \scriptsize
% \tiny

\LinesNumberedHidden 

\textbf{\textit{Initialisation}}\\
\textit{
A stream of pair $(x, y)$, as $(x_{0},y_{0}), (x_{1},y_{1}), \cdots, (x_{T},y_{T})$, arriving
one-by-one over time.
}\\

\textit{X is an evolving data stream (X $\rightarrow$ $\infty$), where $x_{t}$ is  a set of features observed at time $t$ (now).\\}

\textit{$y$ is the ground truth (a.k.a. label) and $\hat{y}$ is the classifier ($h$) prediction\\}

\textit{$m$: maximum features evaluated per split; $n$: total number of Hoeffding trees $(n=|\Psi|)$; $\delta_{w}$: concept-drift detector (ADWIN) warning threshold; $\delta_{d}$: concept-drift detector drift threshold; $c(\cdot)$: change detector (Hoeffding) classifier; $S$: Data stream; B: Set of background Hoeffding trees; $W(t)$: Tree $t$ weight; $P(\cdot)$: estimation function of the learning performance; sgn: sign function (a.k.a signum function); CD: Concept-drift.
\\}
\textit{$x_{i}$ is the i-th input data point} \\
\textit{$L$ is the OCSVMs base learning algorithm} \\
\textit{$K$ is the Number of OCSVMs used in ensemble \scriptsize\ttfamily // Odd number to break the tie} \\

\vspace{1mm}  
\hrule
\vspace{1mm}

 \textit{$f(x)$ $\leftarrow$ sgn(($w.\phi(x_{i})$)-$\rho$)=sgn($\Sigma_{i=1}^{n}$)  \scriptsize\ttfamily  // One Class SVM anomaly detector equation}\\
    \textit{$P(\theta_{*ITA}\geq\theta_{ITA})$ $\leq$ 1-$\alpha$ \scriptsize\ttfamily  // kdqTree concept drift detection equation}

\For(\tcp*[h]{\textit{upon receiving an input ($x_{t}$)}}){each $(x_{t})$ in $X$}
{
   
    \textit{$\delta$ $\leftarrow$ $f(x_{t})$} {\textit{ \scriptsize\ttfamily  // classification of $x_{t}$ by OCSVM}}\\
    \textit{$\delta_{1}$ $\leftarrow$ $P(\theta)$} {\textit{ \scriptsize\ttfamily  // Detects drift or not}}\\

    \If(\tcp*[h]{\textit{$h_{m}$ detect $(x_{t})$ as anomaly}}){$\delta == -1$}
    {
    \textit{$\Psi$ $\leftarrow$ CreateHoeffdingTrees($n$)}\\
    \textit{$W$ $\leftarrow$ InitialiseWeights($n$)}\\
    \textit{$B$ $\leftarrow$ Null}\\
    \textit{$\forall \psi \in \Psi \textbf{do}$:\\}
    
    \hspace*{1em}\parbox[t]{313pt}{
      \textit{$\hat{y}$ $\leftarrow$ ARFTree.predict($\psi$, $x_{t}$)}\; }
    \hspace*{1em}\parbox[t]{313pt}{
      \textit{$W(\psi)$ $\leftarrow$ $P(W(\psi),\hat{y} ,y)$}\; }
      
      \hspace*{1em}\parbox[t]{313pt}{
      \textit{ARFTree.train($m,\psi, x_{t}, y$) \scriptsize\ttfamily  // train model $ \psi$ on ($x_{t}, y$)}\;}\;
      \vspace{-3mm} %Put here to reduce too much white space after your table 
       \hspace*{1.5em}\parbox[t]{313pt}{function ADWIN($\psi$,$x_{t}$)\\
       \hspace*{1.5em} $W$ $\leftarrow$ Initialise Window\\ 
       \hspace*{1.5em} for each $\psi$ $\geq0$ do\\
       \hspace*{2.5em} $W$ $\leftarrow$ $W\cup$ $x_{t}$ {\textit{ \scriptsize \ttfamily  // adding $x_{t}$ to the window}}\\
       \hspace*{2.5em} repeat {\scriptsize \ttfamily // popping elements from the tail of \\ the window}\\
       \hspace*{1.5em} end $\mid$ $W_{0} - W_{1} \mid \ge W_{threshold}$\\
       \hspace*{0.2em} return $ \delta_{w}, \delta_{d}$}\\

      \hspace*{1.5em}\parbox[t]{313pt}{      \If(\tcp*[h]{\textit{CD warning detected?}}){$C(\delta_{w},\psi, x_{t}, y)$} 
      {
         \textit{$b \leftarrow$ CreateHoeffdingTree()  \scriptsize\ttfamily  // Initialise the \\ background ensemble HoeffdingTrees}\;
         \hspace*{-1em}\hspace*{1em}\textit{$B(\psi) \leftarrow b$}\;
         }
      }\;
        \vspace{-3mm} %Put here to reduce too much white space after your table 

    \hspace*{1.5em}\parbox[t]{313pt}{      \If(\tcp*[h]{\textit{CD detected?}}){$C(\delta_{d},\psi, x_{t},y)$}
      {
         \textit{$\psi \leftarrow$ B($\psi$) \scriptsize\ttfamily  // replace $\psi$ by the background \\ ensemble HoeffdingTrees classifier}\;
         }
      }\;
          \vspace{-3mm} %Put here to reduce too much white space after your table 

    \hspace*{.5em}\textit{$\forall b \in B$ $\textbf{do}$: \ \ \scriptsize\ttfamily  //  Train each background HoeffdingTrees}
    \hspace*{1em}\parbox[t]{313pt}{
      \textit{ARFTree.train($m,b, x_{t},y$)}\; }

    % EnsembleOCSVM($x_{t},y,L,K$) \\
    
    \For(\tcp*[h]{generate bootstrap sample and train a base learner on that sample for K different OCSVMs}){k = 1 to K}
    {
     $D_{k}$ = Bootstrap(D) \\
     $h^{OCSVM}_{k}$ = L(D) \\
    %  end \\ 
     }
     \hspace*{1em}{$H(X) = argmax_{y \in Y} \Sigma_{k=1}^{K} l(y = h^{OCSVM}_{k}(x))$} \\
     \textit{\scriptsize\ttfamily //  l(a) = 1: if a is true, 0 otherwise} \\
    
    \If(\tcp*[h]{\textit{$h_{m}$ detect $(x_{t})$ as anomaly}}){$\hat{y} == anomaly$}
    {
    
       \textit{{Trigger the Intrusion Prevention System (As depicted in Fig. \ref{fig:UW-ASN-IPS})}}
       }
    }
    
}
\caption{Proposed Algorithm}
\label{Alg:hybrid-ids}

\end{algorithm}

\section{Implementation and Evaluation}
\label{sec:Implementation and Evaluation}

In this section, we present the results of experiments conducted to evaluate the performance and effectiveness of the proposed scheme.

\subsection{Dataset Generation and Feature Engineering}
\label{sec:Dataset Generation and Feature Engineering}
This paper used the Network Simulator (NS-2) and Aquasim to generate the dataset \cite{das2016simulation}. We considered different network topologies (e.g., 16$\sim$64 nodes, different number of malicious nodes and position of these malicious nodes). Different topologies of the UW-ASN were created, and the vector-based forward routing protocol was used. Table \ref{tab:simulation_parameters} shows the simulation parameters and their values as part of the experiment. The simulation script was then modified to accommodate attacks like blackhole, grayhole and flooding. The blackhole attack was generated by targeting a forwarding sensor node in the network to drop its packets. For the grayhole attack, a selected forwarding node was made to drop a randomly chosen percentage of packets. Three malicious nodes were chosen for the flooding attack to flood a parent node in the network. The execution of the simulation scripts results in the generation of a trace file. Four scripts were used to generate the dataset. The first script (T1) contained a 16-node topology, and the second script (T2) was used for the flooding attack on the 16-node topology. The third script (T3) was used to generate the OOD dataset, which has a 64-node topology, and the fourth script (T4) was used for the flooding attack on the 64-node topology. For the generation of the normal dataset, T1 was used. T1, along with the blackhole attack's algorithm in the vector-based routing script, was used to generate the dataset for the blackhole attack. T1, along with the grayhole attack's algorithm in the vector-based routing script, was used to generate the dataset for the grayhole attack. T2 was used to create the dataset for the flooding attack. The trace files generated were converted to a structured CSV format using regular expressions of Python. Individual datasets were then generated for the normal scenario, blackhole attack, grayhole attack and flooding attack, and merged to form a master dataset. 
 
\begin{table}[htbp]
  \caption{Simulation Parameters}
  \label{tab:simulation_parameters}
  
%   \begin{tabular}{ccl}
\begin{center}
\scalebox{1.}{
  \begin{tabular}{|l|l|}
\hline

    \textbf{Parameters} & \textbf{Values} \\ 
    \hline

    Simulator & Aquasim \cite{xie2009aqua} \\ 
    \hline

    Number of nodes & 16$\sim$64 \\ 
    \hline

    Number of Malicious nodes & 5  \\ 
\hline
    
    Channel & UnderwaterChannel \\ 
\hline
    
    Propagation & UnderwaterPropagation \\ 
\hline
    
    MAC &  BroadcastMac \\ 
\hline

    Initial Energy & 10000 Watt\\ 
    \hline
    
    Antenna & OmniAntenna \\
\hline

    Filters & GradientFilter \\ 
\hline
    
    Max packet in ifq & 50 \\ 
\hline
    
    X dimension of the topography & 100 meters \\ 
\hline
    
    Y dimension of the topography & 100 meters\\ 
\hline
    
    Z dimension of the topography & 100 meters\\ 
\hline
    
    Datarate & 0.1 (1 packet/100 milliseconds)\\ 
\hline
    
    AdhocRouting & Vector Based Forward (VBF) \\ 
\hline
    
    Hop by Hop & 1 \\ %MRHOF, 
\hline
    
    Frequency & 25 kHz \\ 
\hline
    
    Simulation Time & 600 seconds \\ 
\hline

\end{tabular}}
\end{center}
% \vspace{-6mm} %Put here to reduce too much white space after your table 
\end{table}

 Two master datasets,  d1 and d2, were generated. The d1 dataset has the independent variables along with class labels as:
 \begin{itemize}
 \item \textbf{0}: Belonging to the normal class.
 \item \textbf{1}: Belonging to the blackhole class.
 \item \textbf{2}: Belonging to the grayhole class.
 \item \textbf{3}: Belonging to the flooding class.
 \end{itemize}
 
Blackhole was labelled 1 when the receiver was the malicious node(s). Grayhole was labelled 2 when either the sender or receiver was malicious. Flooding was labelled 3 when the sender was a malicious node(s). All other scenarios were labelled 0, i.e., normal. 
 
The d2 dataset has the independent variables along with class labels as:
 \begin{itemize}
 \item \textbf{0}: Belonging to the normal class.
 \item \textbf{1}: Belonging to all the attack classes clubbed together, i.e., blackhole, grayhole and flooding attacks.
 \end{itemize}
 
\begin{figure}[htp]
    \centering
    \includegraphics[width=8cm]{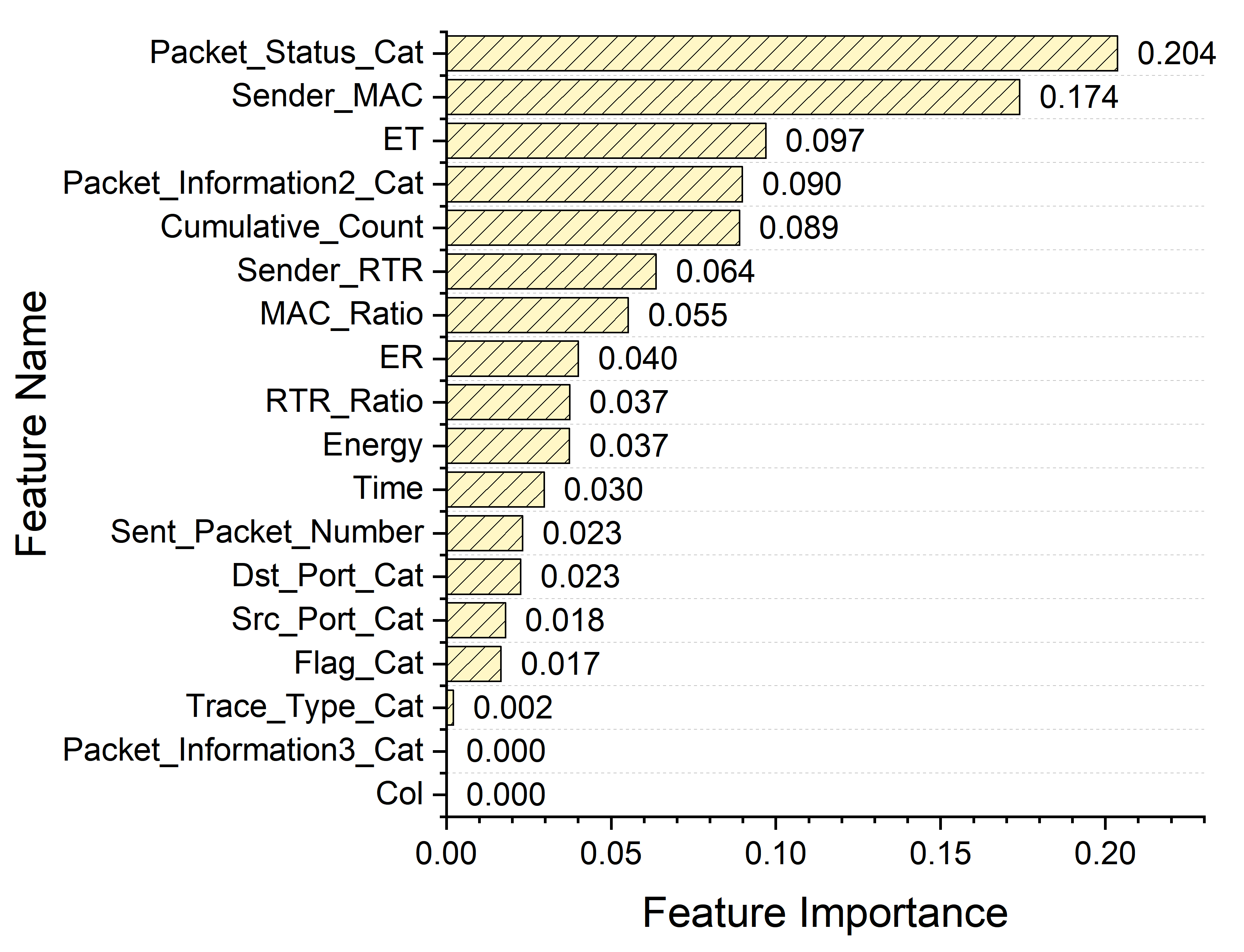}
    \caption{Feature importance by random forest.}
    \label{fig:feature importance}
\end{figure}

As part of the feature engineering process, we first derived five features: Sender RTR (Response Time Reporter), Sender MAC, cumulative count, RTR ratio, and MAC ratio. 
Here, the RTR feature is used to monitor the network performance and resources by measuring response times and the availability of the UW-ASN devices. RTR ratio is the ratio of the Sender RTR to the Cumulative\_Count when the trace type is `RTR'. MAC ratio is the ratio of the Sender MAC to the cumulative count when the trace type is `MAC'. The cumulative count is the incremental count for each trace type, i.e. `RTR' and `MAC'. To check for the overall feature importance, we used Random Forest. Figure \ref{fig:feature importance} shows the outcome of the feature importance given by random forest \cite{jaiswal2017application}. Variation Inflation Factor (VIF) was also used to check for multi-collinearity. Both techniques were used to remove the features that did not add value. Packet Information3 and Col have high VIF values and minor feature importance; thus, these features were removed. The final features are given in Table \ref{tab:Engineered_features}. Our developed dataset contains 29157 instances, 16 independent features, and one dependent feature (a.k.a. target or label).

\begin{table}[t!]
\caption{Engineered Features}
\begin{center}
\begin{tabular}{|c|p{0.55\linewidth}|}
\hline 
\textbf{Feature} & \textbf{Description} \\ 
\hline 
Packet\_Status\_Cat & Categorical column to show the packet category (r: receive, s: send, d: drop) \\ 
\hline 
Sender\_MAC & Numerical column which calculates the Sender MAC value \\ 
\hline
ET & Numerical column which gives the value of ET at different instances. \\ 
\hline
Packet\_Information2\_Cat & Categorical column which contains the application packet's informations. \\ 
\hline 
Cumulative\_Count & Numerical column which calculates the incremental count for each trace type i.e., 'RTR' and 'MAC'. \\ 
\hline
Sender\_RTR & Numerical column which calculates the Sender RTR value  \\ 
\hline
MAC\_Ratio & Numerical column which computes the ratio of the Sender MAC to the Cumulative Count when the trace type is 'MAC'. \\ 
\hline
ER & Numerical column which gives the value of ER at different instances. \\ 
\hline
RTR\_Ratio & Numerical column which computes the ratio of the Sender RTR to the Cumulative Count when the trace type is 'RTR'. \\ 
\hline
Energy & Numerical column which gives the value of the energy at different instances. \\ 
\hline
Time & Time at which the application packet was sent \\ 
\hline
Sent\_Packet\_Number & Numerical column which tells the packet number sent at different instances. \\ 
\hline
Dst\_Port\_Cat & Categorical column which represents the destination port \\ 
\hline
Src\_Port\_Cat & Categorical column which represents the source port \\ 
\hline
Flag\_Cat & Categorical column which descibes the flag type. \\ 
\hline
Trace\_Type\_Cat & Categorical column which descibes the trace type (RTR and MAC)  \\ 
\hline 
Attack\_Cat & Categorical column which represents different scenarios (0: Normal, 1: Blackhole attack, 2: Grayhole attack, 3: Flooding attack) \\ 
\hline 
\end{tabular} 
\label{tab:Engineered_features}
\end{center}
\end{table}

\begin{table*}[t!]
  \caption{Evaluation of different standard ML classification
models against different metrics on training dataset}
  \label{tab:evaluation}
  
%   \begin{tabular}{ccl}
\begin{center}

\scalebox{1.}{
  \begin{tabular}{|c|c|c|c|c|c|c|c|c|}
\hline 
   \textbf{Model} & \textbf{Accuracy} & \textbf{AUC} & \textbf{Recall} & \textbf{Precision} & \textbf{F1-Score} & \textbf{Kappa} & \textbf{MCC} & \textbf{TT (Sec)}\\
  \hline 
  \hline 
  \textbf{Light Gradient Boosting Machine (lightgbm)} & \textbf{0.9997} & \textbf{1.0000}  & \textbf{0.9997} & \textbf{0.9997} & \textbf{0.9997} & \textbf{0.9997} & \textbf{0.9997} & \textbf{0.4180}\\
  \hline 
  Random Forest Classifier (rf) & 0.9994 & 1.0000 & 0.9994 & 0.9994 & 0.9994 & 0.9991 & 0.9991 & 0.6090\\
  \hline 
   Extra Trees Classifier (et) & 0.9993 & 1.0000 & 0.9993 & 0.9993 & 0.9993 & 0.9991 & 0.9991 & 0.2200\\
  \hline 
   Decision Tree Classifier (dt) & 0.9988 & 0.9992 & 0.9988 & 0.9988 & 0.9988 & 0.9984 & 0.9984 & 0.0680\\
  \hline 
   Gradient Boosting Classifier (gbc) & 0.9980 & 1.0000 & 0.9980 & 0.9980 & 0.9980 & 0.9973 & 0.9973 & 5.7930\\
  \hline 
   K Neighbors Classifier (knn) & 0.9912 & 0.9973 & 0.9912 & 0.9914 & 0.9912 & 0.9883 & 0.9884 & 6.8060\\
  \hline 
   Logistic Regression (lr) & 0.9800 & 0.9961 & 0.9800 & 0.9807 & 0.9798 & 0.9733 & 0.9737 & 1.8310\\
  \hline 
   SVM - Linear Kernel (svm) & 0.9725 & 0.0000 & 0.9725 & 0.9737 & 0.9720 & 0.9633 & 0.9639 & 0.0960\\
  \hline 
   Naive Bayes (nb) & 0.9613 & 0.9902 & 0.9613 & 0.9634 & 0.9602 & 0.9484 & 0.9497 & 0.0500\\
  \hline 
   Ridge Classifier (ridge) & 0.8973 & 0.0000 & 0.8973 & 0.9091 & 0.8945 & 0.8630 & 0.8681 & 0.0270\\
  \hline 
   Linear Discriminant Analysis (lda) & 0.8953 & 0.9802 & 0.8953 & 0.9074 & 0.8923 & 0.8604 & 0.8657 & 0.0470\\
  \hline 
   Ada Boost Classifier (ada) & 0.7225 & 0.9344 & 0.7225 & 0.6554 & 0.6571 & 0.6301 & 0.6734 & 0.3900\\
  \hline 
   Quadratic Discriminant Analysis (qda) & 0.2500 & 0.0000 & 0.2500 & 0.0625 & 0.1000 & 0.0000 & 0.0000 & 0.0290\\
  \hline 
   Dummy Classifier (dummy) & 0.2499 & 0.5000 & 0.2499 & 0.0625 & 0.0999 & 0.0000 & 0.0000 & 0.0430\\
  \hline    
\end{tabular}}

\end{center}
% \vspace{-6mm} %Put here to reduce too much white space after your table 

\end{table*}

\subsection{Performance Evaluation and Discussion}
\label{sec:Performance_Evaluation}

To measure the effectiveness of the proposed scheme, we conducted eleven experiments to justify the contributions and to evaluate the extent to which the proposed scheme is capable of addressing each of the desirable properties discussed in Section \ref{subsec:Desirable_Properties}.

\subsubsection{\textbf{Experiment 1 (Anomaly Detectors)}}

As discussed in Section \ref{sec:Anomaly-based IDS}, we deploy anomaly-based IDS on the underwater monitoring sensor nodes to detect anomalous network communications. In this regard, we benchmark the performance of the proposed scheme under different one-class classifiers (a.k.a. outlier detectors), such as OCSVM, LOF and isolation forest (IF):

\begin{itemize}

 \item \textbf{One-class support vector machine (OCSVM):} OCSVM is a semi-supervised (it trains on normal behaviours) anomaly or outlier detector which creates a decision boundary, inside which it classifies data points as normal or inlier and outside which it classifies the data points as abnormal or outlier. For hyper-parameter tuning of OCSVM, we took different values of $\gamma$ ranging from 0.1 to 0.5 and different values of $\nu$ ranging from 0.004 to 0.05. $\gamma$ decides how much curvature is needed in a decision boundary. The parameter $\nu$ fine tunes the trade-off between overfitting and generalization. The optimal parameter value for $\nu$ is 0.01 (meaning that at most 1\% of the training samples are outliers by the decision boundary) and for $\gamma$ is 0.3 \cite{eude2018one}.

 \item \textbf{Local outlier factor (LOF):} LOF is also a semi-supervised anomaly/outlier detector and can detect novelties in the dataset. It compares the local density of each data point to that of its neighbours. Data points with higher densities are classified as normal, whereas those with lower densities are classified as anomalies or outliers. For hyper-parameter tuning of LOF, we used different values for contamination ranging from 0.0001 to 0.5.

 \item \textbf{Isolation forest (IF):} IF is also a semi-supervised anomaly detection algorithm similar to random forest as it is built using decision trees. It works on the principle that data points that are easy to separate by the trees are considered anomalies. In contrast, the data points that were relatively difficult to separate are normal. It considers a random subset of the data and a random subset of the features. For hyper-parameter tuning of IF, we used different values for contamination ranging from 0.0001 to 0.5. The contamination value is float, and should be in the range (0, 0.5] \cite{hastuti2020designing}. The contamination parameter simply controls the threshold for the decision function when a scored data point should be considered an outlier.
 
 \end{itemize}

 We also evaluated our proposed framework under Adaptive One-Class Support Vector Machine (AOCSVM) and QuantileFilter. AOCSVM performed well in true positive and false negative ratios, scoring 0.959 and 0.04, respectively. However, it had low true negative and high false positive ratios. The evaluation result of QuantileFilter was almost comparable to that of AOCSVM.
 OCSVM outperforms all the other anomaly detectors with the kernel: RBF, $\nu= 0.01$ and $\gamma= 0.3$. The AIDS using OCSVM was evaluated, and it showed an accuracy of 0.9374, recall or true positive ratio (TPR) of 1.0, F1-score of 0.9662, AUC of 0.9374, false negative ratio (FNR) of 0.0, true negative ratio (TNR) of 0.9374, precision of 0.935 and false positive ratio (FPR) of 0.0626. 
%Figure 1 shows the decision boundary created by OCSVM, the anomaly data points and the data points considered normal by the OCSVM anomaly/outlier detection algorithm. 
Though LOF and IF show good TPR and FNR, they have low TNR and FPR. OCSVM, on the other hand, shows promising results across all the metrics. The specific network characteristics, data patterns, and the nature of the underwater environment in the UW-ASN influence the performance of the different anomaly detectors.

To decentralise the proposed scheme, we suggest placing the AIDS in specific underwater sensor nodes, preferably on parent sensor nodes. Figure \ref{fig:UW-ASN} shows the decentralised architecture of the proposed scheme. HIDS in the figure represents the hybrid intrusion detection system, a combination of anomaly and signature-based IDS. Our proposed solution combines OCSVM for AIDS and adaptive random forest for SIDS. HIDS gets the global view of the network and is placed at the surface station (the surface buoy). AIDS passively monitors (a.k.a. monitoring in promiscuous mode) its neighbouring nodes' network communications. Although our proposed scheme is network-based and performs in silent mode (a.k.a. ghost mode), to avoid the single point of failure, we consider a decentralised monitoring and intrusion detection approach by distributing the AIDS agents in the UWASN.

\subsubsection{\textbf{Experiment 2 (Incremental Machine Learning Classifiers)}} As discussed in Section \ref{sec:Proposed Incremetanl IDS}, our proposed scheme adopts an Incremental Machine Learning classifier to classify observations in evolving data streams. Hence, the proposed IDS can adapt incrementally.
In this regard, our proposed scheme uses ARF for this task. The ARF algorithm (an ensemble of Hoeffding trees) introduces diversity through re-sampling and randomly selecting a subset of features on the streaming data environment. Moreover, using a concept drift detector (discussed in Section \ref{sec:preliminaries_concept_drift}), it replaces weak learners (a.k.a. losers) with the new learners (Hoeffding trees) constructed on recent concept drifts. 

The evaluation of signature-based IDS using the ARF model shows an accuracy of 0.9774, a precision of 1.0, an AUC of 0.9887, a true negative ratio of 1.0, and a false positive ratio of 0.0. The recall and the true positive ratio value were 0.9775 and the false negative ratio is 0.0225. Factors that contributed to this experiment are the incremental machine learning algorithm and the selection of appropriate parameters. In addition to these, diversity and feature selection also played a crucial role in driving the results.

\subsubsection{\textbf{Experiment 3 (Concept Drift Detection)}}
\begin{figure}[htp]
    \centering
    \includegraphics[width=9cm]{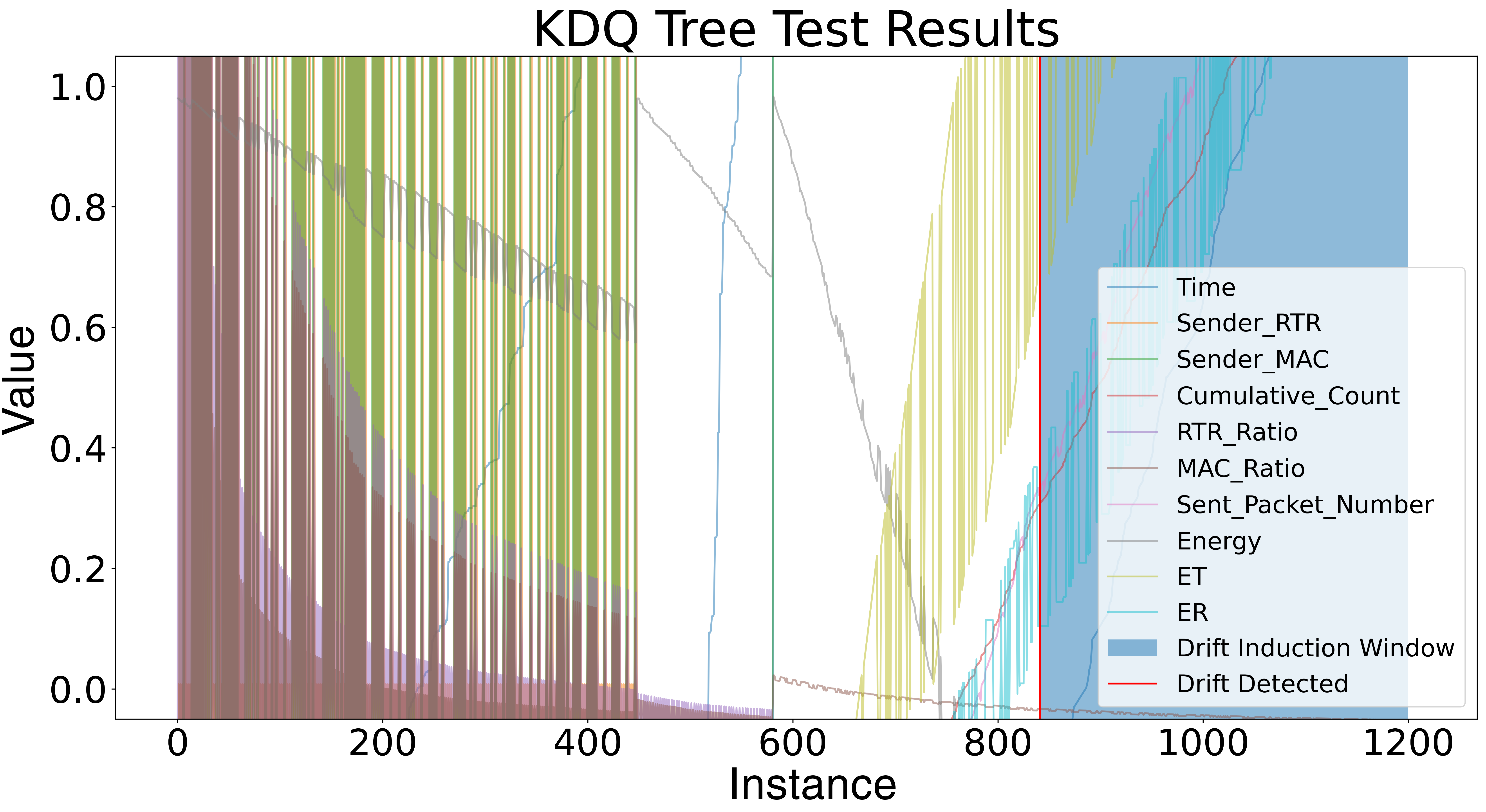}
    \caption{kdqTree concept drift detection.}
    \label{fig:kdqTree}
\end{figure}

As discussed in Section \ref{sec:Proposed Incremetanl IDS}, the proposed scheme also adopts unsupervised and supervised concept drift detectors (kdqTree and ADWIN, respectively). In this regard, the proposed scheme implements a data-distribution-based central unsupervised concept drift detection algorithm for its solution. kdqTree has been implemented to understand when, how, and where the drift has occurred. It internally uses KL Divergence to detect the drifts. This information is then fed to the ARF algorithm, which uses a supervised error-rate-based concept drift warning detector and drift signal detector, i.e., ADWIN. ADWIN detects when the drift has occurred and uses this information to update the ARF, which has been trained on a new random subset of data and features. We evaluated ARF with different supervised concept drift detection algorithms like ADWIN, DDM, EDDM, HDDM\_A, HDDM\_W, Kolmogorov-Smirnov Windowing (KSWIN) and Page-Hinkley \cite{lu2018learning}. ARF and ADWIN ($\delta=0.001$) outperformed the other concept drift detection algorithms. For the kdqTree, we used a window size of 500, $\alpha = 0.05$, a bootstrap sample of 500 and a count bound value of 50 \cite{long2011statistical}. Figure \ref{fig:kdqTree} shows the visualisation of the drift detected by kdqTree on the UW-ASN streaming data.
%The figure shows the instance at which drift is detected by kdqTree (marked as a red line on the x-axis). 
The figure plots the streaming data instances along the x-axis with respect to the other variables in the UW-ASN dataset along the y-axis. A drift is detected by kdqTree at instance 841 (marked as a red line on the x-axis) and the blue window thereafter represents the drift induction window.

Figure C.14 and Fig. C.15 of Appendix C in the Supplementary Material show the concept drift detected in the blackhole and flooding attacks, respectively, using ADWIN. Though ADWIN was not successful in detecting the grayhole attack, it was noticed by other concept drift detection algorithms like HDDM\_A, KSWIN and PageHinkley. Fig. C.16, Fig. C.17 and Fig. C.18 from Appendix C of the Supplementary Material show the concept drift detected in the grayhole attack using HDDM\_A, KSWIN and PageHinkley. Implementing a data-distribution-based concept drift detector was important as it is relatively expensive to have labels all the time to use a supervised concept drift detection algorithm, especially to recognise a zero-day (unseen) intrusion related to UW-ASNs. This eliminates the expectation of having a label available as soon as the new data has arrived. For unsupervised concept drift, we focus on the probability distribution of the feature variables $P(X)$ and compare any quantifying change in this distribution. To detect concept drifts on the distribution of the data stream, we adopt kdqTree using a sliding window in stage 1. Next, kdqTree employs Kulldorff's spatial scan statistic to identify the regions with the most changes in the streaming data. At  the final stages, kdqTree uses the KL divergence test and implements bootstrapping method \cite{lu2018learning}. The Kullback-Leibler Divergence score, also known as KL divergence score, measures the disparity between two probability distributions. Overall, the selection of appropriate concept drift detection algorithms and the utilization of data-distribution-based detection techniques contributed to the successful identification of concept drifts in this experimental setup.

\subsubsection{\textbf{Experiment 4 (Proposed Hybrid Solution)}}

In this experiment, we evaluate the overall performance of the proposed scheme.
The network-based IDS, which implements AIDS on the parent sensor nodes, implements the OCSVM anomaly detection algorithm using RBF kernel, and $\nu$ and $\gamma$ are assigned to 0.01 and 0.3, respectively (based on the outcome of Experiment 1). OCSVM model is evaluated against the testing dataset, and the evaluation results are shown in Table \ref{tab:evaluation_testing}. Adaptive Random Forest (ARF), along with kdqTree, is able to detect 28 drift instances in the UWASN dataset. This information was further fed to the SIDS. The SIDS employs ADWIN concept-drift detector with drift warning  and drift detection parameters both assigned as $\delta=0.001$. ARF employs 50 estimators (Hoeffding trees), maximum features to consider at each node, for each time to make the split decision as log2 and split confidence as 0.001. The SIDS model is evaluated against the testing dataset, and the evaluation results are also shown in Table \ref{tab:evaluation_testing}. The final predictions are taken from the hybrid model and evaluated against the underwater testing dataset. Our proposed hybrid model has an accuracy of 0.9774, precision of 0.9717, recall of 1.0, F1-score of 0.9883, AUC of 0.977, TPR of 1.0, FNR of 0.0, TNR of 0.9769 and FPR of 0.023. The results are shown in Table \ref{tab:evaluation_testing}.

\begin{table*}[t!]
  \caption{Evaluation Results on Testing Dataset}
  \label{tab:evaluation_testing}
  
%   \begin{tabular}{ccl}
\begin{center}

\scalebox{1.2}{
  \begin{tabular}{|c|c|c|c|c|c|c|c|c|c|}
\hline 
  \textbf{Model} & \textbf{Precision} & \textbf{Accuracy} & \textbf{Recall} & \textbf{F1-score} & \textbf{AUC} & \textbf{TPR} & \textbf{FNR} & \textbf{TNR} & \textbf{FPR}\\
  \hline 
  \hline  
 OCSVM & 0.935 & 0.9374 & 1.0 & 0.9662 & 0.9374 & 1.0 & 0.0 & 0.9347 & 0.0626\\
  \hline
 ARF and kdqTree & 1.0 & 0.9774 & 0.9775 & 0.9883 & 0.9887 & 0.9775 & 0.0225 & 1.0 & 0.0\\
  \hline
 \textbf{Proposed Hybrid Model} & \textbf{0.9717} & \textbf{0.9774} & \textbf{1.0} & \textbf{0.9883} & \textbf{0.977} & \textbf{1.0} & \textbf{0.0} & \textbf{0.9769} & \textbf{0.023}\\
  \hline

\multicolumn{10}{l}{$^{\mathrm{*}}$TPR: True Positive Ratio. FNR: False Negative Ratio. TNR: True Negative Ratio. FPR: False Positive Ratio. 
}
\\
\multicolumn{10}{l}{$^{\mathrm{}}$OCSVM: One Class Support vector Machine. ARF: Adaptive Random Forest. 
}
\\
\end{tabular}}

\end{center}
% \vspace{-6mm} %Put here to reduce too much white space after your table 

\end{table*}

Throughout our experiment, we focused on improving the true positive ratio and reducing the false negative ratio. We also focused on reducing the false positive ratio to less than 5\%, which helps in reducing false alerts or false alarms to a great extent. Several factors influenced the results of this experiment. Firstly, the performance of the OCSVM anomaly detection algorithm in AIDS played a crucial role in the overall results. Secondly, the effectiveness of the drift detection mechanism implemented by ARF and kdqTree was crucial in adapting to evolving data streams. Lastly, by optimising the performance metrics like true positive ratio and false negative ratio, the proposed hybrid solution demonstrated improved performance in accurately identifying anomalies while reducing false alarms.

\subsubsection{\textbf{Experiment 5 (Optimisation of ARF classification algorithm)}}
\begin{figure}[htbp]
    \centering
    \includegraphics[width=9cm]{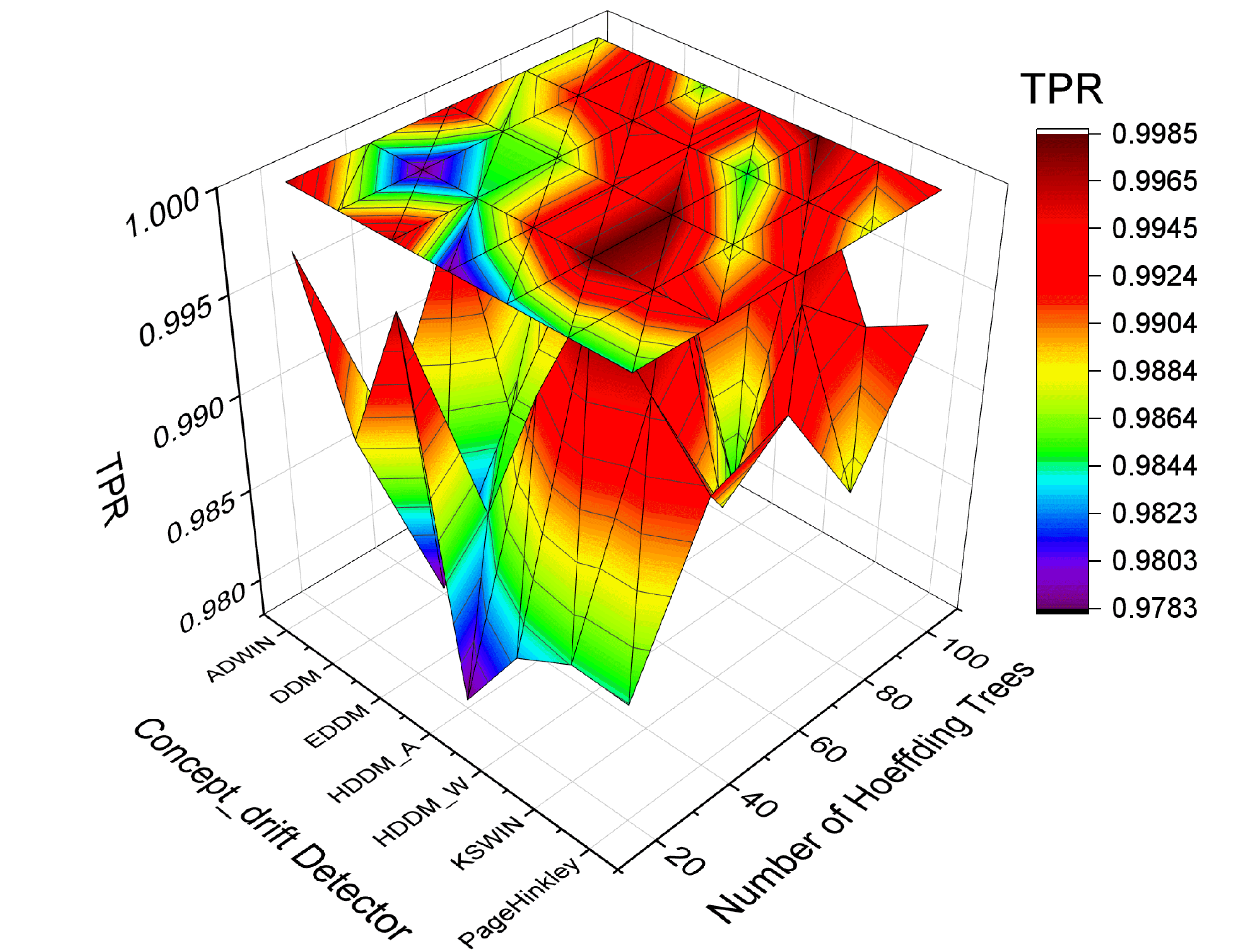}
    \caption{Optimisation analysis for the proposed scheme.}
    \label{fig:optimization}
\end{figure}

In this experiment, we optimise the performance of the ARF classifier. To receive an optimal set of parameter values, we trained the ARF classification algorithm with different numbers of Hoeffding trees (20, 40, 60, 80 and 100). We used different concept drift detection algorithms (ADWIN ($\delta=0.001$), DDM, EDDM, HDDM\_A, HDDM\_W, KSWIN and PageHinkley) along with it, keeping all the other configurations from our proposed scheme constant. We then evaluated the algorithm's performance on the TPR metric. Figure \ref{fig:optimization} shows the visualisation from our analysis. The three-dimensional graph shows the ensemble of Hoeffding trees, different concept drift detectors used and the true positive ratio for each. This experiment gives an idea of the optimal number of estimators (number of ensemble Hoeffding trees) to make the ARF classification algorithm perform effectively when coupled with any of the above-mentioned concept drift detectors. For dealing with an evolving streaming data environment, it becomes essential to make the model optimal to reduce the computational complexity without compromising its performance.  The graph depicts the optimal parameters (concept drift detector and the number of trees) that we need for our proposed scheme to reduce the computational overhead as well as maintain the TPR. By exploring the relationship between the number of Hoeffding trees, concept drift detectors, and the TPR metric, the experiment provides insights into the optimal configuration of the ARF classifier. These factors contribute to reducing computational overhead while maintaining high performance in intrusion detection tasks.

TPR is used as it is an important metric that calculates how accurately our proposed system can detect an intrusion event. ARF optimisation was also performed with other metrics and has been provided in the Appendix E of the Supplementary Material.

\subsubsection{\textbf{Experiment 6 (Scalability with 64 sensor nodes)}}
In this experiment, to evaluate the scalability of the proposed scheme, we generate a scenario with 64 acoustic sensor nodes where 20\% were malicious nodes. In this scenario, malicious nodes conduct blackhole, grayhole, and flooding attacks against their neighbouring nodes.
In this context, UWASN nodes' distribution,  distance, and velocity are different from our proposed scheme development phase (e.g., where we trained our AIDS on normal network communication). Therefore, we now evaluate our proposed scheme against the OOD data stream. The derived OOD dataset has 67614 instances along with the same 16 features and one target variable.

After the final OOD dataset was generated, it was passed through the proposed scheme to evaluate the results. Our proposed hybrid model has an accuracy of 0.99, precision of 0.99, recall of 1.0, F1-score of 0.995, AUC of 0.99, TPR of 1.0, FNR of 0.0, TNR of 0.99 and FPR of 0.01. The OCSVM and SIDS model's evaluation results on the OOD dataset are given in Table \ref{tab:ood}. ARF and kdqTree detected 65 drift instances in the underwater OOD dataset. The proposed hybrid model gave the final predictions and was evaluated on the underwater OOD dataset. The results from this experiment prove the scalability of our proposed solution. As UW-ASNs can utilise large numbers of underwater wireless sensor nodes connected to work together, making the proposed IDS solution scale well to achieve specific tasks is essential. This, in turn, makes the proposed solution achieve DP4.

Several factors influenced the results obtained in this experiment. Firstly, for these evaluation results, the OOD dataset derived from a different distribution and representing scenarios not encountered during the development phase was used to assess the system's performance in recognizing unseen intrusions and adapting to new situations. Secondly, the hybrid model that combines multiple algorithms (OCSVM, ARF and kdqTree) is another key factor driving the results, allowing the system to effectively handle underwater intrusion detection tasks in dynamic environments. The OCSVM, ARF and kdqTree model cannot operate as standalone models as they won’t be able to satisfy all Desirable Properties (for instance ARF alone fails to recognise unseen intrusions on-the-fly as it is signature-based, and OCSVM is not adaptive as its decision boundary is static). The proposed model combines the strength of all these algorithms and works together in tandem  to achieve all the desirable properties listed in section 
\ref{subsec:Desirable_Properties}.

\begin{table*}[t!]
  \caption{Evaluation Results on OOD Dataset}
  \label{tab:ood}
  
%   \begin{tabular}{ccl}
\begin{center}

\scalebox{1.2}{
  \begin{tabular}{|c|c|c|c|c|c|c|c|c|c|}
\hline 
  \textbf{Model} & \textbf{Precision} & \textbf{Accuracy} & \textbf{Recall} & \textbf{F1-score} & \textbf{AUC} & \textbf{TPR} & \textbf{FNR} & \textbf{TNR} & \textbf{FPR}\\
  \hline 
  \hline  
 OCSVM & 0.9144 & 0.9154 & 1.0 & 0.9553 & 0.9146 & 1.0 & 0.0 & 0.9145 & 0.0854\\
  \hline
 ARF and kdqTree & 1.0 & 0.9901 & 0.9138 & 0.995 & 0.9569 & 0.9138 & 0.0862 & 1.0 & 0.0\\
  \hline
 \textbf{Proposed Hybrid Model} & \textbf{0.99} & \textbf{0.99} & \textbf{1.0} & \textbf{0.995} & \textbf{0.99} & \textbf{1.0} & \textbf{0.0} & \textbf{0.99} & \textbf{0.01}\\
  \hline
\multicolumn{10}{l}{$^{\mathrm{*}}$TPR: True Positive Ratio. FNR: False Negative Ratio. TNR: True Negative Ratio. FPR: False Positive Ratio. 
}
\\
\multicolumn{10}{l}{$^{\mathrm{}}$OCSVM: One Class Support vector Machine. ARF: Adaptive Random Forest. OOD: Out-of-Distribution
}
\\

\end{tabular}}

\end{center}
% \vspace{-6mm} %Put here to reduce too much white space after your table 

\end{table*}

\subsubsection{\textbf{Experiment 7 (Generalisability)}}

In this experiment, we generate different types of concept drifts (e.g., abrupt, recurring, gradual and incremental) to measure up to what extent our proposed system can identify and adapt to concept drifts \cite{gama2014survey} and how well the concept drift detectors generalise on the new data stream. ADWIN ($\delta=0.001$), DDM, HDDM\_A, KSWIN and PageHinkley were used to detect the drifts. Different drift detectors were able to capture drifts in various instances. Figure A.1 of Appendix A of the Supplementary Material shows the drifts and the cases as and when ADWIN detected them. Figure A.2 from Appendix A of the Supplementary Material shows the drifts seen by DDM. The error rate decreases as the samples used for analysing increase, as long as the data distribution remains constant. Figure A.3, Fig. A.4 and Fig. A.5 of Appendix A of the Supplementary Material show the drifts detected by HDDM\_A, KSWIN and PageHinkley, respectively. The results of this experiment show that the concept drift detectors used as part of our proposed scheme could also be leveraged to be used in other streaming datasets, thereby making them reusable for any application area dealing with streaming data environments. The choice of concept drift detection algorithms and the analysis from the visualizations provided were key factors driving the results, highlighting the system's ability to detect and handle drifts in streaming data environments.

\subsubsection{\textbf{Experiment 8 (Benchmarking standard ML classifiers)}}
In this experiment, we benchmarked the IDS by applying it against standard machine learning classification algorithms and evaluating them against the metrics: accuracy, AUC, recall, precision, F1-score, kappa and MCC. The list of standard ML classifiers and their results for different metrics are provided in Table \ref{tab:evaluation}.

Light Gradient Boosting Machine (LGBM) outperformed all the other standard ML classifiers on all the above mentioned metrics. LGBM had an accuracy of 0.99954, AUC of 0.99843, recall of 0.99723, a precision of 0.991758, F1-score of 0.99976, a true positive ratio of 0.99723, a false negative ratio of 0.00276, a true negative ratio of 0.99964 and a false positive ratio of 0.00035 on the testing dataset (refer to Table \ref{tab:evaluation}). LGBM was used as the SIDS.

%We evaluated OCSVM, Local Outlier Factor (LOF) and Isolation Forest for the anomaly-based detection system.  The parameters used by OCSVM are \emp{$\nu$}, \emp{$\gamma$} and \emp{kernel}. The parameter $\nu$ is used to specify the percentage of anomalies. The kernel is used to identify the kernel type and also maps the data to a higher dimensional space for the SVM to draw a decision boundary. The parameter $\gamma$ is used to set the kernel coefficient. OCSVM has the best performance with the kernel, $\nu$, and $\gamma$ assigned as RBF, 0.01, and  0.3, respectively. 
%OCSVM is used as the anomaly-based detector and gave an accuracy of 0.9515, AUC of 0.9747, recall of 1.0, a precision of 0.4605, F1-score of 0.97406, a true positive ratio of 1.0, a false negative ratio of 0.0, a true negative ratio of 0.9494 and a false positive ratio of 0.0505 on the testing dataset.
OCSVM is used as the anomaly-based detector and gave an accuracy of 0.9374, AUC of 0.9374, recall of 1.0, a precision of 0.935, F1-score of 0.9662, a true positive ratio of 1.0, a false negative ratio of 0.0, a true negative ratio of 0.9347, and a false positive ratio of 0.0626 on the testing dataset.

For the benchmarked model, final predictions were taken from the OCSVM when LGBM predicted the instance as a regular instance, else from LGBM when LGBM predicted the instance as an attack. The final predictions from the hybrid model were then evaluated against the ground truth. The IDS used for benchmarking had an accuracy of 0.9514, AUC of 0.9746, recall of 1.0, a precision of 0.4599, F1-score of 0.9740, a true positive ratio of 1.0, a false negative ratio of 0.0, a true negative ratio of 0.9493 and a false positive ratio of 0.0506 on the testing dataset.

In this experiment, we use the same features that were engineered in Section \ref{sec:Dataset Generation and Feature Engineering} (to have a fair comparison between our proposed scheme and the benchmarked algorithms). We employ an Auto-ML library (Pycaret) to automate the hyperparameter tuning process. It implements 10-fold cross-validation to tune the LightGBM model by finding the best set of parameter values. In this regard, the final optimised LightGBM model has a learning\_rate of 0.1, max\_depth of -1, num\_leaves as 31, and n\_estimators as 100. The detailed configurations of the remaining benchmarked classifiers from Table \ref{tab:evaluation} are provided in Appendix F of the Supplementary Material.

%boosting\_type = 'gbdt', colsample\_bytree = 1.0, importance\_type = 'split', learning\_rate = 0.1, max\_depth = -1, min\_child\_samples = 20, min\_child\_weight = 0.001, min\_split\_gain = 0.0, n\_estimators = 100, n\_jobs = -1, num\_leaves = 31, reg\_alpha = 0.0, reg\_lamda = 0.0, silent = 'warn', subsample = 1.0, subsample\_for\_bin = 200000 and subsample\_freq = 0. The configuration of OCSVM model is kept same as that of the proposed scheme for a fairer comparison. The parameters and theirs values for OCSVM are: kernel = 'rbf', nu = 0.01 and gamma = 0.3.

By comparing the performance of standard machine learning classifiers and selecting LightGBM as the top-performing classifier, the experiment demonstrated the effectiveness of the proposed IDS. The hyperparameter tuning process and the limitations of the benchmarked classifiers were factors that influenced the obtained results. Benchmarked classifiers in Tables \ref{tab:evaluation} are not adaptive and do not scale well. In addition, they fail to recognise unseen (zero-day) intrusions on-the-fly. Hence, they require continuous monitoring of the model's performance, which is expensive. 
A graphical representation of Table \ref{tab:evaluation}, Table \ref{tab:evaluation_testing}, and Table \ref{tab:ood} is presented in Appendix G of the Supplementary Material.

\subsubsection{\textbf{Experiment 9 (Model comparison with respect to concept drift)}}
\begin{figure}[htp]
    \centering
    \includegraphics[width=9cm]{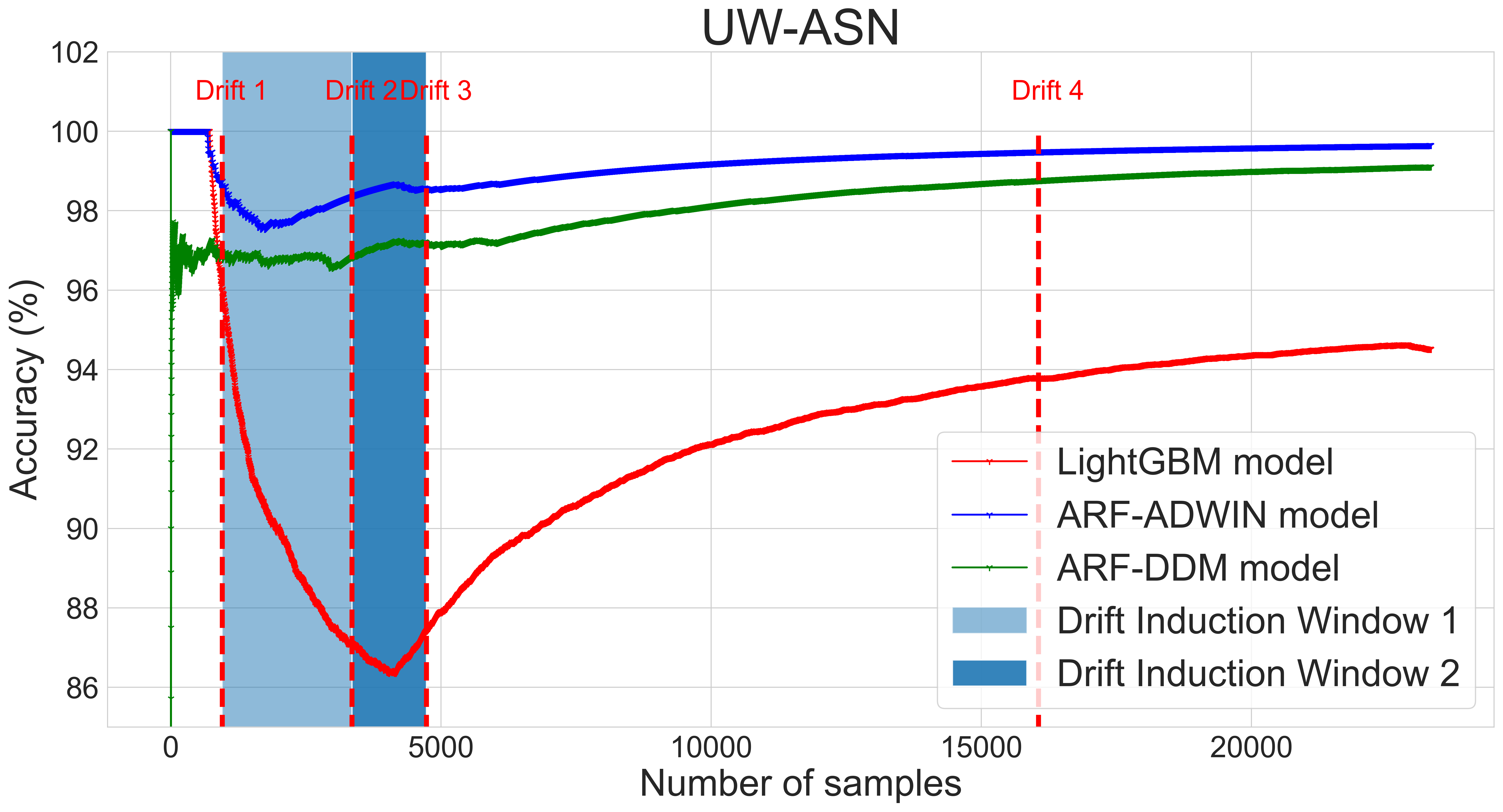}
    \caption{Signature-based IDS model comparison against concept drifts.}
    \label{fig:drifts}
\end{figure}

In this experiment, we compare signature-based IDS models like Light GBM (used for benchmarking), ARF with ADWIN (used for the proposed scheme), and ARF with DDM. The three IDSs were evaluated over the accuracy metric to analyse their performance against concept drifts.
Figure \ref{fig:drifts} plots the streaming data samples along the x-axis with respect to the accuracy along the y-axis. The continuous red line, green line, and the blue line plots the accuracy of the Light GBM model, ARF with DDM concept drift detection algorithm, and ARF with ADWIN (used in the proposed scheme), respectively. The graph also plots four drift instances marked with red dotted lines along the x-axis. This graph shows that the proposed IDS model (ARF with ADWIN) maintains consistent accuracy even during the concept drifts as opposed to the other two IDS models. Factors driving the results of this experiment are adaptability to concept drift, the choice of drift detection algorithms, and the design of the IDS models.

\begin{figure*}[t!]
\begin{subfigure}{.5\textwidth}
\centerline{\includegraphics[width=7.5cm]{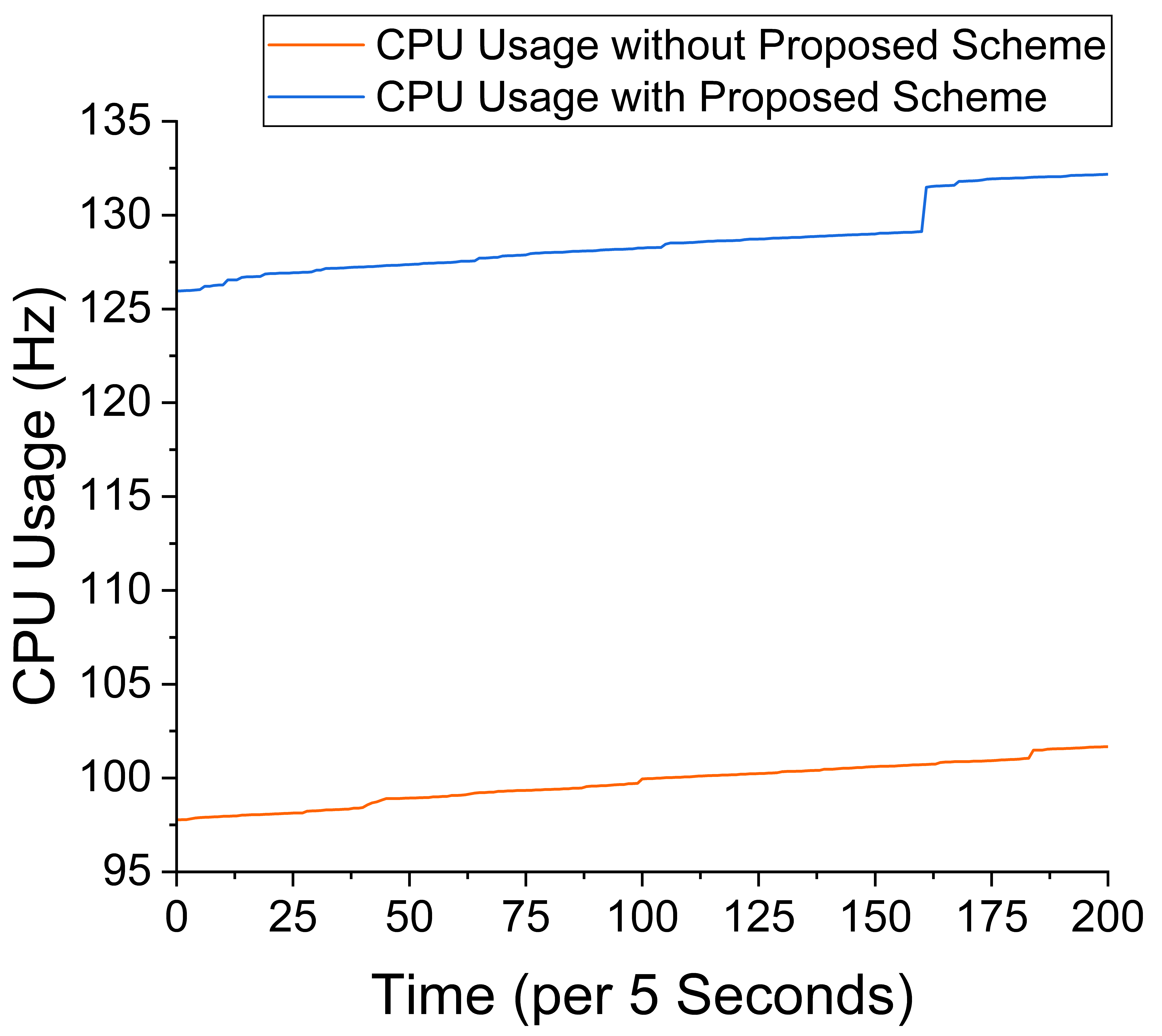}}
\caption{CPU usage of a sensor node (Raspberry Pi 3B+)}
% \label{fig:CPU_Usage}
\end{subfigure}%
\begin{subfigure}{.5\textwidth}
\centerline{\includegraphics[width=8.5cm]{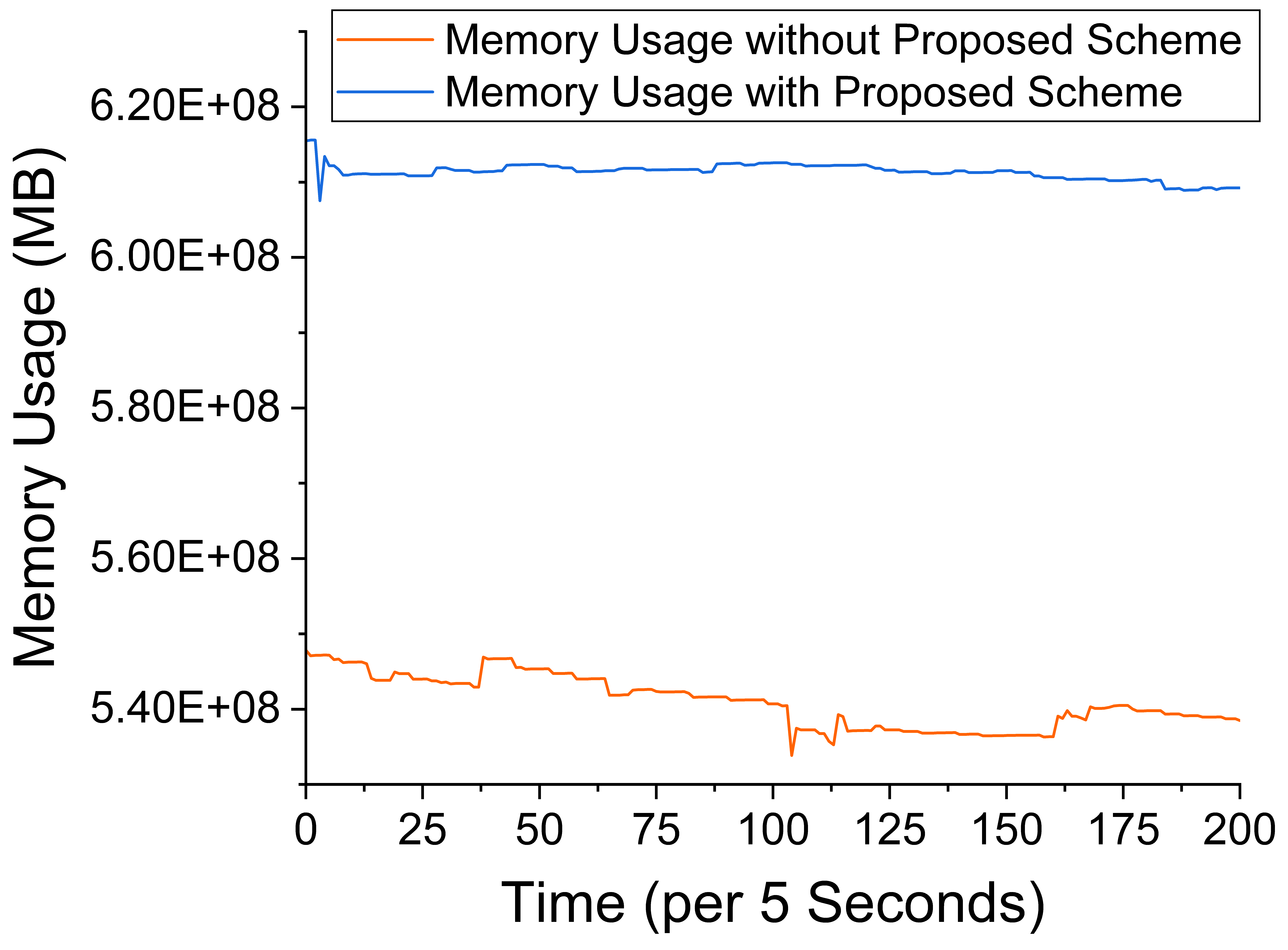}}
\caption{Memory usage of a sensor node (Raspberry Pi 3B+)}
% \label{fig:Memory_Usage}
\end{subfigure}
\caption{CPU and Memory usages of a sensor node (Raspberry Pi 3B+)}
\label{fig:CPU and Memory Usage}
% \vspace{-6mm} %Put here to reduce too much white space after your table 
\end{figure*}

\subsubsection{\textbf{Experiment 10 (Exploratory data analysis)}}

In this experiment, we perform exploratory data analysis (EDA) to show the statistical analysis of all the numerical features with respect to the different types of attacks. In this regard, as part of the EDA process, we investigated more on the numerical features. Data distributions were compared by calculating the $p$-value, and outlier detection was visualised with the help of box plots. The graphs are shown in Appendix B of the Supplementary Material in Fig. B.6, Fig. B.7, Fig. B.8, Fig. B.9, Fig. B.10, Fig. B.11, Fig. B.12 and Fig. B.13 for ET, Sender\_MAC, Energy, Sender\_RTR, Sent\_Packet\_Number,
RTR\_Ratio, MAC\_Ratio and ER, respectively. 
The factors driving the results of this experiment are as follows. The experiment aimed to perform exploratory data analysis (EDA) on numerical features in order to gain insights into their statistical characteristics in relation to different types of attacks. This involved a thorough investigation of the numerical features, comparing their data distributions through $p$-value calculations, and visualizing outlier detection using box plots. By conducting this EDA, a deeper understanding of the numerical features and their associations with different attack types was achieved.

\subsubsection{\textbf{Experiment 11 (Energy Consumption)}}
\begin{figure}[htp]
    \centering
    \includegraphics[width=8.4cm]{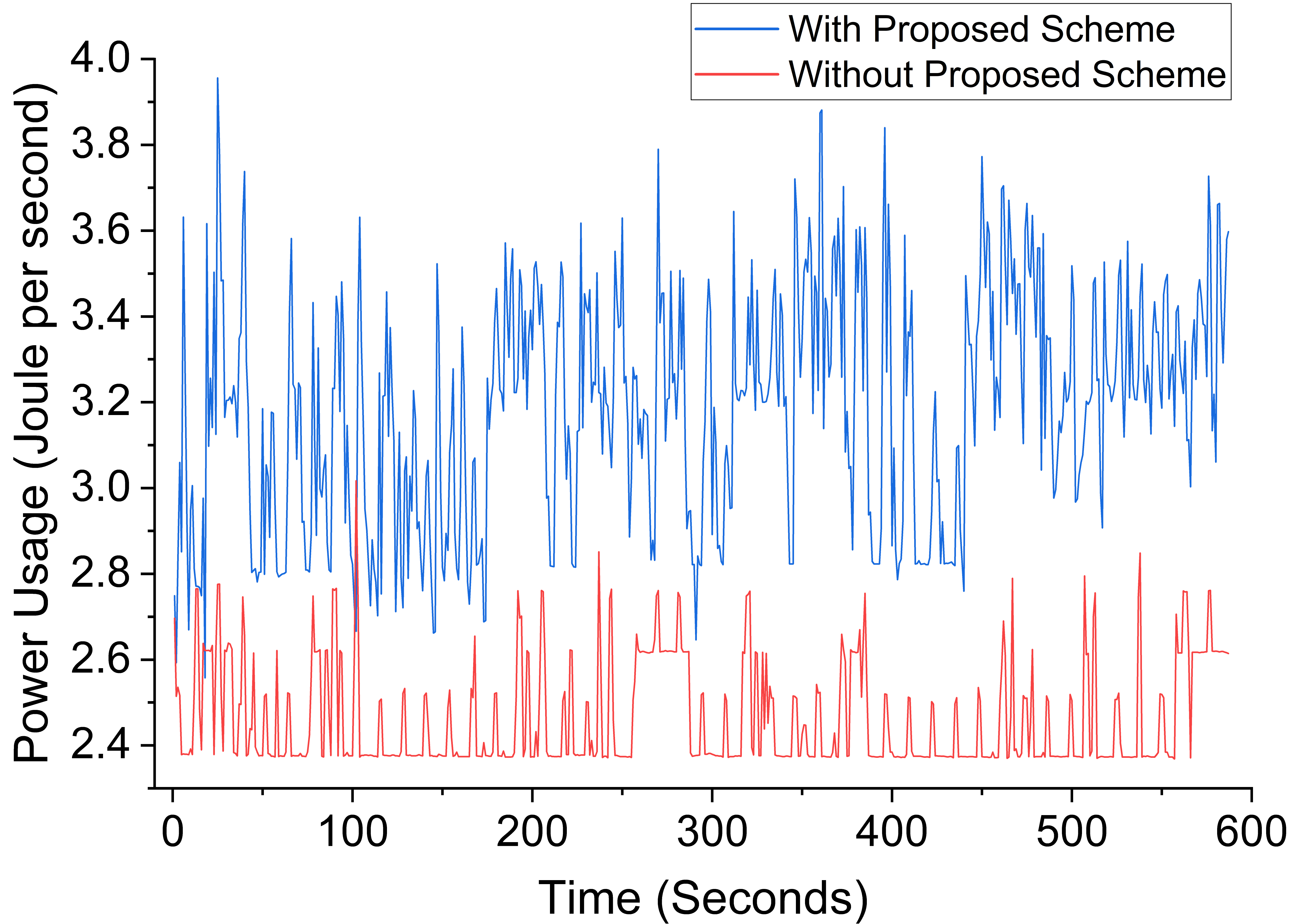}
    \caption{Energy Overhead}
    \label{fig:Power Usage}
\end{figure}

Since the UW-ASN contains resource constrained sensor nodes, our proposed scheme has to be lightweight  (\emph{DP6}). 
The Raspberry Pi 3B+ microprocessor has been considered a resource-constrained sensor node. The pre-trained anomaly detector was deployed using a socket program in the microprocessor to make the sensor have lower computational overhead. The socket program at the client side (the central edge computer) sent each application packet one at a time to the server side (the microprocessor representing the parent sensor node enabled with listening mode). The microprocessor utilised the pre-trained anomaly detector to pass each packet through the model to get the prediction on-the-fly. UM25C was connected to the microprocessor to analyse the power consumption. The power consumption is captured over time (separately when the socket program was not operational and when it was active). A side-channel analysis was performed (shown in Fig. \ref{fig:Power Usage}) to analyse the energy overhead caused by the anomaly detector.

The average power usage (in Joule per second) when the algorithm was not operational for the microprocessor was 2.46 J/s, and when the algorithm was active was 3.18 J/s. The difference in energy overhead is 0.72 J/s which provides evidence for the proposed scheme having the desirable property DP6.

The CPU and memory usage for the microprocessor were logged every five seconds when the proposed scheme was not operational and when it is operational. The data is used to analyse and calculate the CPU and memory usage overhead. 
%The graphs are provided in Appendix D of the Supplementary Material in Fig. D.19 and Fig. D.20.

Figure \ref{fig:CPU and Memory Usage}(a) shows the CPU usage comparison of a sensor node (Raspberry Pi 3B+) when it was operational with and without the proposed scheme. The CPU usage was logged every 5 seconds. The average CPU usage (in Hz) particularly when the proposed scheme is not operational for the microprocessor is 100.02 Hz, and when the proposed scheme is operational was 128.88 Hz. The difference in CPU usage overhead is 28.86 Hz, which provides evidence for the proposed scheme having the desirable property DP6.

Figure \ref{fig:CPU and Memory Usage}(b) shows the memory usage comparison of a sensor node (Raspberry Pi 3B+) when it was operational with and without the proposed scheme. The memory usage was logged every 5 seconds. 

The factors driving the results of this experiment are as follows. Firstly, the proposed scheme was designed to be lightweight, addressing the resource constraints of UW-ASNs. The Raspberry Pi 3B+ microprocessor was chosen as a suitable resource-constrained sensor node. Secondly, the deployment of a pre-trained anomaly detector using a socket program minimized computational overhead, allowing real-time predictions. Thirdly, power consumption analysis using UM25C demonstrated a small energy overhead of 0.72 J/s, supporting the scheme's lightweight nature. Additionally, monitoring of CPU usage revealed a modest increase in usage (28.86 Hz) during operation. Overall, these results indicate that the proposed scheme satisfies DP6 of the desirable properties.

The outcomes of our experiment show that the proposed scheme is capable of satisfying all the desirable properties listed in Section \ref{subsec:Desirable_Properties}. This makes the proposed scheme efficient in detecting zero-day intrusions and scale well with the increase in the number of underwater sensor nodes, and is lightweight to work in a resource-constrained environment.

\section{Conclusion and Future Works}
\label{sec:Conclusion}
UW-ASNs find applications in many areas, and their threats are significant. Therefore, in this paper, we proposed an adaptive, hybrid, distributed IDS that is efficient in handling evolving underwater streaming data and detecting the changes in the underlying data patterns, and a cryptography-based IPS as an autonomous self-defence mechanism. We have also integrated several concepts like incremental machine learning and data distribution-based concept drift detection algorithms, which can be leveraged to be used in other application domains dealing with streaming data environments. An attack dataset for UW-ASN-based IoT networks, which covers three attack types (blackhole, grayhole and flooding attacks) specific to UW-ASN, is introduced. The dataset (UW-ASN dataset) and the proposed IDS scheme can be benchmarked to be used for UW-ASNs-related works by the research community.
%Our evaluation results show that the proposed scheme outperforms state-of-the-art benchmarking methods while providing a wider range of desirable features (DP1 to DP7). \\
%\textcolor{blue}{Our proposed scheme integrating incremental machine learning and concept drift detection algorithms can not only autonomously detect intrusion events in an evolving underwater streaming data environment but also, removes the pain and cost of continuously monitoring the machine learning model for its performance. In addition to this, the cryptography-based IPS autonomously helps in mitigating intrusions to prevent the adversary from taking further actions.}\\
%As part of the future works related to the security of UW-ASNs, possible areas for extending the proposed work include:
The proposed scheme outperforms state-of-the-art benchmarking methods while providing a wider range of desirable features (DP1 to DP7). While our proposed scheme used a fixed set of rules for the anomaly detectors, we recommend enhancing its intelligence by generating dynamic rules and giving them priority. In this context, we propose the following future works related to the security of UW-ASNs:

\begin{itemize}

\item Low-Rate DoS attack: Low-Rate DoS attacks are an intelligent adaptation of DoS attacks where the malicious nodes flood the parent sensor nodes with application packets. However, the frequency of messages is kept below the approved threshold level to make it difficult for an IDS to detect such attacks \cite{zhijun2020low}. 
\item Deep Adversarial Reinforcement Learning (ARL) based IDS: Deep ARL for a signature-based IDS helps to generate a dynamic set of rules and prioritise them specifically for a given environment. This technique can enhance the performance of a SIDS. 
\item Data distribution-based concept drift detection: Enhance the performance of kdqTree to work efficiently with Incremental Machine Learning algorithms.

\end{itemize}

\bibliographystyle{IEEEtran} 
\bibliography{Main_Manuscript}

\vskip -2\baselineskip plus -1fil

\begin{IEEEbiography}
[{\includegraphics[width=1.2in,height=1.35in,keepaspectratio]{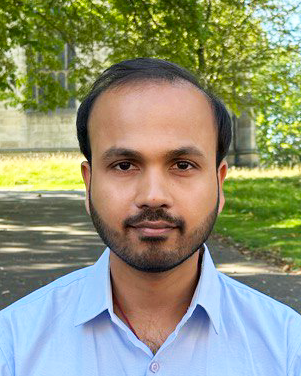}}]{Soumadeep Das}

is a researcher with the Security of Advanced Systems Research Group at the University of Sheffield. He earned his Master's degree in Cybersecurity and Artificial Intelligence in 2022. With over six years of industry experience, Soumadeep has excelled as an Artificial Intelligence engineer and has an established record in Data Engineering and Data Science domains. His primary research areas include Intrusion Detection, Reinforcement Learning, and the Penetration Testing of IoT (Internet of Things) networks and embedded devices.

\end{IEEEbiography}
% \vskip 0pt plus -1fil
\vskip -2.5\baselineskip plus -1fil

\begin{IEEEbiography}
[{\includegraphics[width=1.2in,height=1.35in,keepaspectratio]{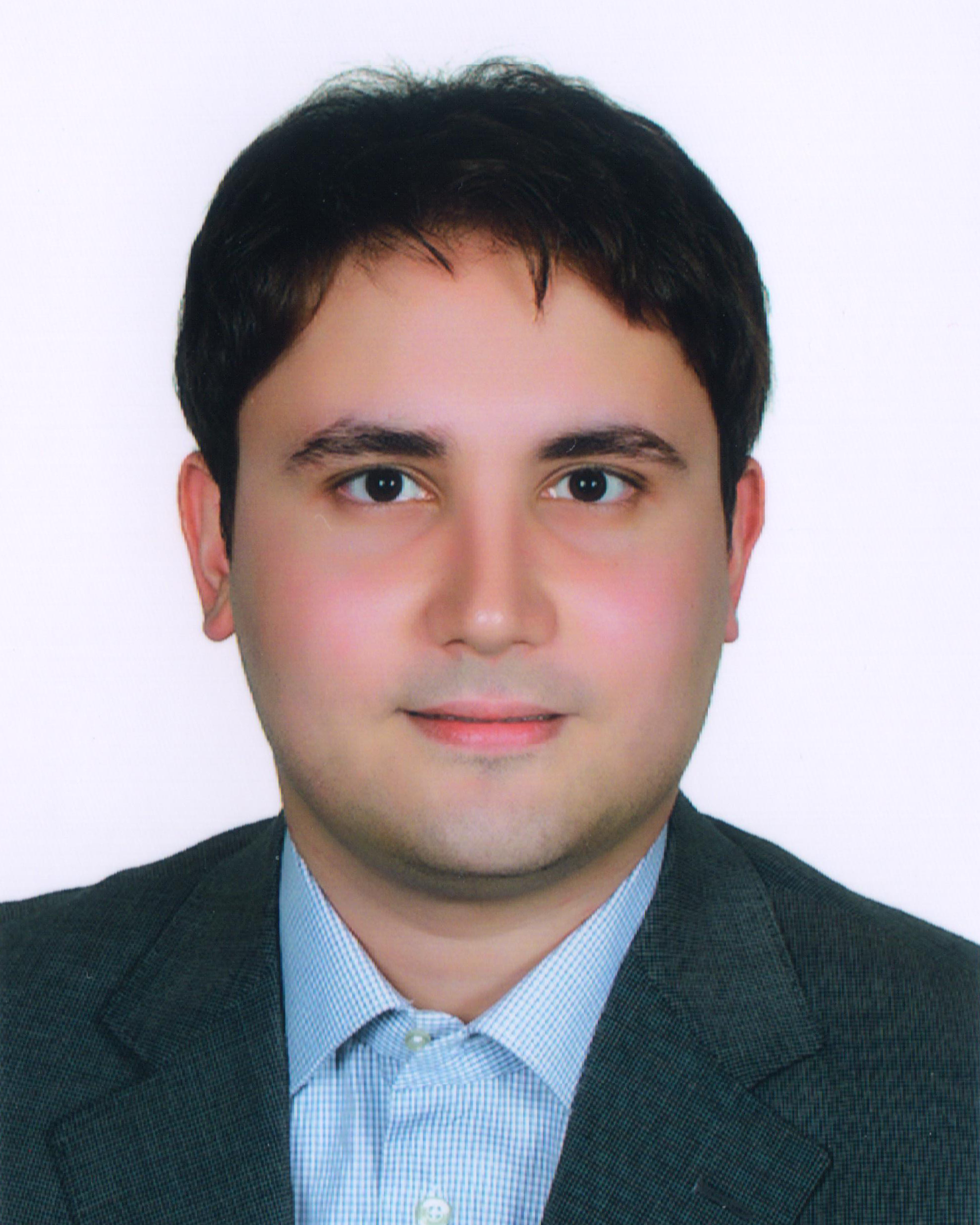}}]{Aryan Mohammadi Pasikhani}

has been a Post-Doctoral Researcher with the Security of Advanced Systems Group at the University of Sheffield. He is currently an Academic Fellow in Cybersecurity at the same institution. His focus is on publishing research that makes a high impact in the fields of Computer and Network Security. His primary research interests encompass intrusion detection and prevention systems, reinforcement learning, machine learning, quantum computing, and the security of embedded systems. Furthermore, he has been an esteemed Technical Program Committee (TPC) Member and has undertaken peer review responsibilities for several prestigious international journals and conferences, including the IEEE Internet of Things Journal, IEEE Transactions on Industrial Informatics, and the IEEE Sensors Journal.

\end{IEEEbiography}
% \vskip 0pt plus -1fil
\vskip -2.5\baselineskip plus -1fil

\begin{IEEEbiography}[{\includegraphics[width=1.2in,height=1.35in,keepaspectratio]{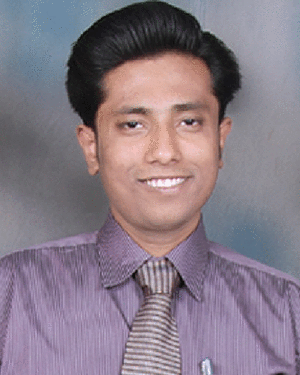}}]{Prosanta Gope}

(Senior Member, IEEE) Currently working as an Assistant Professor with the Department of Computer Science (Cyber Security), University of Sheffield, U.K. He was a Research Fellow with the Department of Computer Science, National University of Singapore. He has authored more than 100 peer-reviewed articles in several reputable international journals and conferences and has four filed patents. Several of his papers have been published in high-impact journals (such as IEEE Transactions on Information Forensics and Security, IEEE Transactions on Dependable and Secure  Computing) and prominent security conferences (such as IEEE Euro S\&P, IEEE  IEEE Computer Security Foundations Symposium (CSF), and IEEE Host). Primarily driven by tackling challenging real-world security problems, he has expertise in lightweight anonymous authentication, authenticated encryption, access control, security of mobile communications, healthcare, the Internet of Things, Cloud, RFIDs, WSNs, smart-grid, and hardware security of the IoT devices. He has served as the TPC Member/Chair in several reputable international conferences, such as ESORICS, IEEE TrustCom,  ARES, etc. He is an Associate Editor for the IEEE Internet of  Things Journal, IEEE Systems Journal, IEEE Sensors Journal, and the Journal of Information Security and Applications (Elsevier).

\end{IEEEbiography}

\vskip -2.5\baselineskip plus -1fil

\begin{IEEEbiography}[{\includegraphics[width=1.2in,height=1.35in,keepaspectratio]{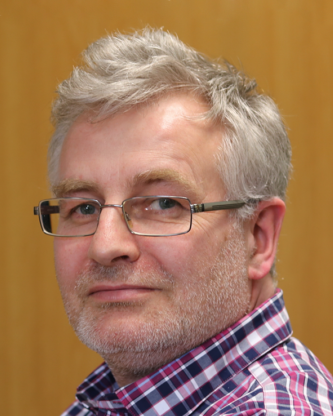}}]{John A Clark}

is the Professor of Computer and Information Security at the University of Sheffield and leads the Security of Advanced Systems Research Group. Previously he was Professor of Critical Systems at the University of York. His major research interests lie in cybersecurity and software engineering, most notably the use of Artificial Intelligence to these areas. Publications have included work on: threat modelling, security policies, covert channel analysis, cryptographic building blocks, intrusion detection, insider detection, and automated synthesis of security protocols. Current work addresses the automated discovery of classical cryptanalytic strategies, intrusion detection (particularly in IoT systems) and its optimal configuration, and consent in healthcare IoT systems.

\end{IEEEbiography}

% % \vskip 0pt plus -1fil
% \vskip -4\baselineskip plus -1fil

% \begin{IEEEbiography}[{\includegraphics[width=1.2in,height=1.35in,keepaspectratio]{Authors/chintan.png}}]{Chintan Patel}

% (Member, IEEE) is a research scholar with the School of Technology, Pandit Deendayal Petroleum University, Gandhinagar, India. With the active academician, he is an active researcher also. He has published papers in various reputed journals like the Wireless Personal Communication (Springer). Recently, he has authored a book titled “ Internet of Things Security Challenges, Advances, and Analytics” with CRC Press, Taylor and Francis Group. He is an active member of the ACMand CSI.

% \end{IEEEbiography}

\vskip -4\baselineskip plus -1fil

\begin{IEEEbiography}[{\includegraphics[width=1.2in,height=1.35in,keepaspectratio]{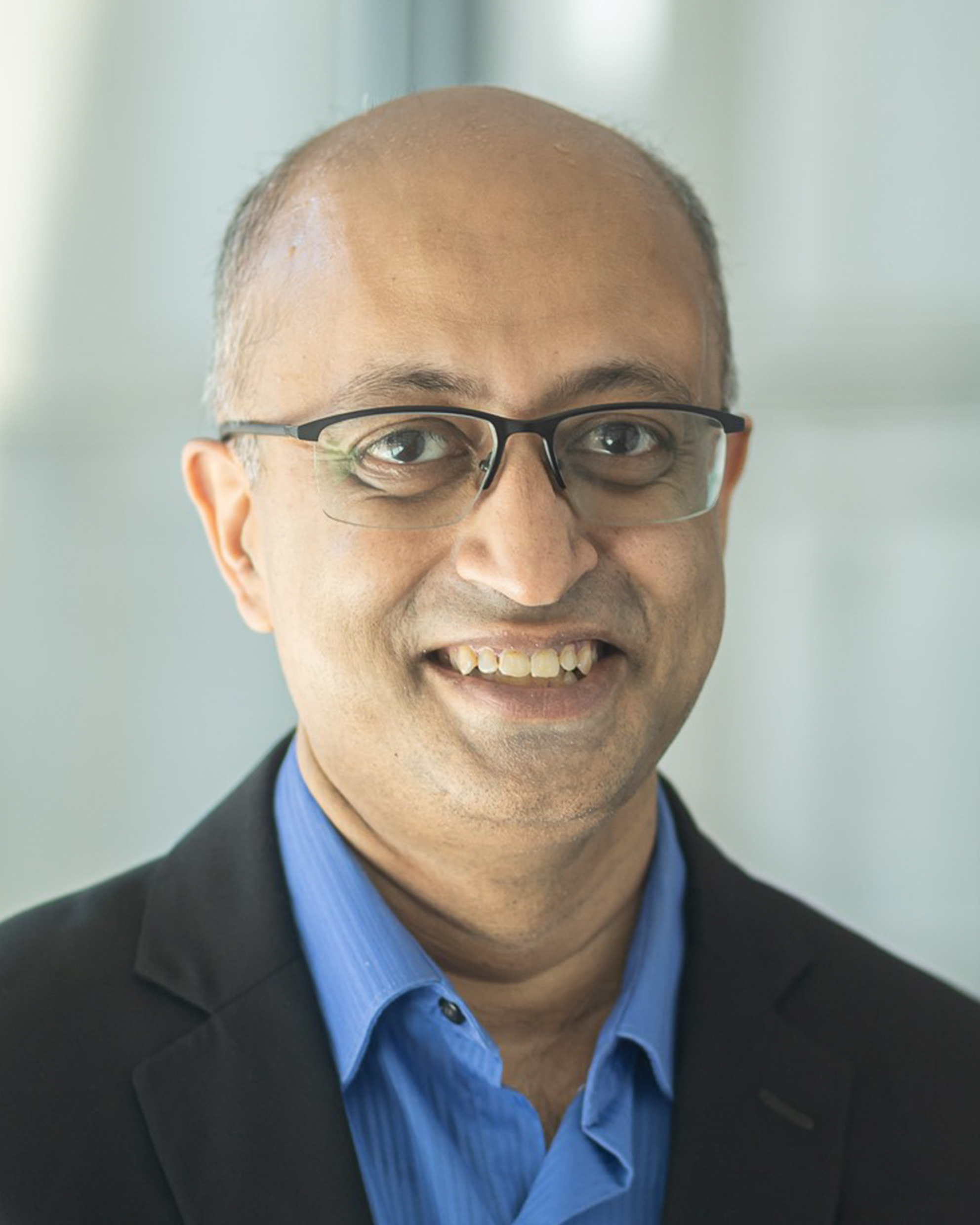}}]{Biplab Sikdar}

(Senior Member, IEEE) received the B.Tech. Degree in electronics and communication engineering from North Eastern Hill University, Shillong, India, in 1996, the M.Tech. Degree in electrical engineering from the Indian Institute of Technology Kanpur, Kanpur, India, in 1998, and a Ph.D. in electrical engineering from the Rensselaer Polytechnic Institute, Troy, NY, USA, in 2001. He was a Faculty with the Rensselaer Polytechnic Institute, from 2001 to 2013, an Assistant Professor and an Associate Professor. He is a Professor and Head of the Department of Electrical and Computer Engineering, at the National University of Singapore, Singapore. He also serves as the Director of the Cisco-NUS Corporate Research Laboratory. His research interests include wireless networks and security for the Internet of Things and cyber-physical systems. He has served as an Associate Editor for the IEEE Transactions on Communications, IEEE Transactions on Mobile Computing, IEEE Internet of Things Journal and IEEE Open Journal of Vehicular Technology.

\end{IEEEbiography}

\end{document}